\documentclass[preprint]{aastex631}
\usepackage{epsfig}
\usepackage{graphicx}
\usepackage{upgreek}
\usepackage{array}
\usepackage{longtable}
\usepackage{amssymb}
\usepackage{amsmath}
\usepackage[fleqn]{mathtools}
\usepackage{mathrsfs}
\usepackage{xcolor}
\usepackage{physics}
\usepackage{bm}
\usepackage{outlines}

\newcommand{\beq}{\begin{equation}}
\newcommand{\eeq}{\end{equation}}
\newcommand{\bea}{\begin{eqnarray}}
\newcommand{\eea}{\end{eqnarray}}
\newcommand{\bean}{\begin{eqnarray*}}
\newcommand{\eean}{\end{eqnarray*}}

\newcommand{\sSun}{ {\scalebox{0.8}{$\odot$}} }

\newcommand{\bB}{ \mathbf{B} }

\newcommand{\bu}{ \mathbf{u} }

\newcommand{\sA}{ {\scalebox{0.6}{\rm A}} }

\newcommand{\sC}{ {\scalebox{0.6}{\rm C}} }

\newcommand{\sH}{ {\scalebox{0.6}{\rm H}} }
\newcommand{\sI}{ {\scalebox{0.6}{\rm I}} }

\newcommand{\sR}{ {\scalebox{0.6}{\rm R}} }
\newcommand{\sS}{ {\scalebox{0.6}{\rm S}} }

\newcommand{\er}{ \hat{\mathbf{e}}_\sR }

\newcommand{\divg}{ \grad \!\cdot \!}

\newcommand{\ie}{{i.e., }}

\newcommand{\eg}{{e.g., }}
\newcommand{\alf}{{Alfv{\'e}n}}

\renewcommand{\arraystretch}{1}
\makeatletter
\let\overlinewithoriginalheight\overline
\newcommand*\overlinewithlessheight[1]{{\mathpalette\overline@aux{#1}}}
\newcommand*\overline@aux[2]{
  \begingroup
    \count0=\fam 
    \setbox0=\hbox{$\m@th #1\fam=\count0 #2$}
    \@tempdima=.4\ht0
    \setbox0=\hbox{$\m@th #1\fam=\count0\overlinewithoriginalheight{#2}$}%
    \advance\@tempdima by .6\ht0
    \ht0=\@tempdima 
    \usebox0
  \endgroup
}
\let\overline\overlinewithlessheight
\let\underlinewithoriginaldepth\underline
\newcommand*\underlinewithlessdepth[1]{{\mathpalette\underline@aux{#1}}}
\newcommand*\underline@aux[2]{%
  \begingroup
    \count0=\fam
    \setbox0=\hbox{$\m@th #1\fam=\count0 #2$}%
    \@tempdima=.4\dp0%
    \setbox0=\hbox{$\m@th #1\fam=\count0\underlinewithoriginaldepth{#2}$}%
    \advance\@tempdima by .6\dp0%
    \dp0=\@tempdima
    \usebox0%
  \endgroup%
}
\let\underline\underlinewithlessdepth
\makeatother
\makeatletter
\allowdisplaybreaks[1]
\usepackage{hyperref}
\graphicspath{{./}{Figs/}}
\submitjournal{\apj}
\shorttitle{Simulating Tomorrow's Solar Wind}
\shortauthors{Sokolov \& Gombosi}
\begin{document}
\title{Physics-Based Forecasting of Tomorrow's Solar Wind at 1 AU}
\correspondingauthor{Igor Sokolov}
\email{igorsok@umich.edu}
\author[0000-0002-6118-0469]{Igor V. Sokolov}
\author[0000-0001-9360-4951]{Tamas I. Gombosi}
\affiliation{University of Michigan, Department of Climate and Space, Ann Arbor, MI, USA}
\begin{abstract}
Inspired by the concept of \textit{relativity of simultaneity} used in the theory of special relativity, a new approach is proposed to simulate \textit{future} solar wind conditions at any point in the inner solar system.  An important distinctive feature of the proposed approach is that the simulation in the solar corona is driven by hourly updated solar magnetograms and is continuously simulated in nearly real time. The model for the inner heliosphere is based on time transformation to a boosted spacetime 
coordinate system, in which the current state of the solar wind at the solar corona -- inner heliosphere boundary and future states of the solar wind are simultaneous. The predictive capability for tomorrow's parameters of the ambient solar wind at 1 AU is achieved by simulating them simultaneously with the current observations of the solar magnetic field, the time offset being enabled by the use of boosted frame.

We derive the modified governing equations for both hydrodynamics and magnetohydrodynamics and present a new numerical algorithm that solves the modified governing equations. The proposed method enables an efficient numerical implementation and thus a significantly longer forecast time than traditional solution methods. In the numerical test for transient propagation, the boosted solution for the CME-driven shock arrival at 1AU is 16 hours ahead of the solution at the solar corona -- inner heliosphere boundary.
\end{abstract}

\keywords{Solar wind (1534) --- Space weather (2037) --- Magnetohydrodynamical simulations (1966) --- Solar-terrestrial interactions (1473)}

\section{Introduction}
\label{Sec:Introduction}
\subsection{Mathematical Motivation}
\label{Subsec:Motivation}
Yogi Berra, the legendary baseball player and philosopher, is credited with saying: ``It is tough to make predictions, especially about the future.'' In modern times, data-assimilative physics-based models revolutionized short-term weather forecasts. These models combine physics models with real-time data from satellites, radars, and weather stations to provide detailed and reliable forecasts for up to three days. Unfortunately, we do not have similar advances in space weather forecasting due to the sparsity of data sources, the vast volume of space needed to be modeled for 24h to 72h forecasting, the lack of advanced data assimilation technologies incorporating sparse observations to model forecasts, and the absence of high-quality parameterizations of subgrid physical processes.

A new way to forecast solar wind is driven by the concept of  ``\textit{relativity of simultaneity}.'' For example, consider a ``Lorentz boosted'' or ``boosted'' frame of reference, moving with a relativistic velocity of $\Lambda$ parallel to the $x$ axis in three-dimensional (3D) Cartesian coordinates, $\mathbf{x} = (x, y, z)$ with respect to the rest frame. Herewith, and throughout this paper, the term ``frame'' combines the 4D \textit{spacetime} coordinates, $\mathbf{x}^{(4)} = (t, \mathbf{x})$ and the transformation between frames, including the transformation of the time component, $t$. 

Let us denote the spacetime coordinates in the boosted frame by subscript $b$, to distinguish them from those in the rest frame (without subscript). Now, the transformation between the two systems, the \textit{Lorenz boost}, is given by 
\begin{eqnarray}
\label{eq:lorentz1}
\renewcommand{\arraystretch}{1.25}
t &=& \Gamma\left(t_b + \frac{\Lambda}{c^2} x_b\right) \\
x &=& \Gamma \left(x_b + \Lambda t_b\right) \\
y &=& y_b \\
z &=& z_b
\renewcommand{\arraystretch}{1.}
\end{eqnarray}
where $c>\Lambda$ is the limiting speed of the transformation (in special relativity, this is the speed of light) and $\Gamma = 1/\sqrt{1-\Lambda^2/c^2}$ is the Lorentz factor. Consider two events that occur simultaneously (or are ``synchronous'', a term often used in the literature) in the boosted frame at the instant of time $t_b^{(1)}=t_b^{(2)}=0$, but at two different locations, $x_b^{(1)}=0$ and $x_b^{(2)}>0$. From the time transformation equations one can immediately see that in the rest frame the events are not simultaneous; they occur at time instants $t^{(1)}=t^{(1)}_b=0$ and $t^{(2)}= \Gamma \Lambda(x_b^{(2)}-x_b^{(1)})/c^2>t^{(2)}_b$. The farther apart the two locations are (increasing $\Delta x_b = x_b^{(2)}-x_b^{(1)}$), the further ahead the rest frame time at the second location ($t^{(2)}$) of the boosted frame time ($t^{(2)}_b$). In Eq.~(\ref{eq:lorentz1}) the speed of light represents the fastest speed allowed by Einstein's special relativity: the spacetime transformation given by Eq.~(\ref{eq:lorentz1}) can  be mathematically defined as long as $c$ is larger than the fastest speed in the system.

Although this analogy is far from perfect, the idea of relativity of simultaneity inspired us to explore a new mathematical framework of a boosted frame, in which the time $t_b$, coincides with the observation time of the latest solar magnetogram, as well as with the measurement time of the ``future'' solar wind at 1 AU. Naturally, the time transformation is parameterized by some speed, which, similarly to the speed of light in Einstein's special relativity, must exceed the maximum perturbation speed for the mathematical validity of the approach.

\subsection{Tomorrow's Space Weather}
\label{Subsec:tomorrow}
The present generation of physics-based operational space weather models has limited prediction capabilities. The most widely used solar wind model, WSA-Enlil integrates an empirical model connecting solar magnetograms to solar wind speeds at around 20 $R_\sSun$ \citep{arge2011,arge2013} and a 3D MHD model of the inner heliosphere \citep{Odstrcil03, Odstrcil99a, Odstrcil99b}. This model is used by NOAA's Space Weather Forecast Office to provide a few days of advance warning of solar wind structures and Earth-directed coronal mass ejections (CMEs). Although widely used, WSA-Enlil has several limitations, including simplistic inner boundary conditions for the interplanetary magnetic field and plasma density and the absence of magnetically driven CME initiation.

The operational SWMF/Geospace model \citep{Toth:2005swmf, Gombosi:2021rev} is driven by \textit{in situ} observations at the L1 point, and thus it only provides up to an hour forecast of the state of Earth's magnetosphere. This forecast window is too short for practical use: by the time the warning of an upcoming geomagnetic storm is issued, the forecast has already expired. There is great need for a new space weather forecast system that starts at the Sun, provides a few days of forecast time at L1, and runs the Geospace model fast enough to provide at least a couple of days of warning of upcoming geomagnetic storms.

This paper describes a ``Sun2Mud'' forecast model that can achieve this objective. A new method for solving the governing equations of the solar wind and interplanetary magnetic field is proposed based on time shifting between the solar corona and inner heliosphere domains. These domains already exist in the SWMF/AWSoM suite \citep{Sokolov2013,vanderHolst:2014a,Sokolov2021}, and we propose only some modifications to the governing equations to allow for a significant increase in the forecast time.

\section{Time-shift Between the Corona and the Solar Wind}
\label{Sec:shift}
In the proposed approach, the 3D MHD model of the coupled solar corona (SC) and inner heliosphere (IH) will run regularly to simulate the evolution for an hour of physical time. Every hour, we start a new simulation run from the previous final state,  obtained with the last hourly magnetogram (taken an hour before) with this latest magnetogram as the time-dependent boundary condition at the solar surface. By the end of each simulation session, the solution in the SC is advanced until the time of next magnetogram. With reasonable computational resources, a computation time of less than one hour can be achieved, so that the recurrently simulated state in the SC lags only about an hour behind physical time.

Forecast capability is achieved with a special ``boosted'' frame in the IH computational domain ($R\ge R_{\sS\sC/\sI\sH}$, $R=\|\mathbf{x}\|$ is the heliocentric distance, $R_{\sS\sC/\sI\sH}$ is the radius of the boundary between SC and IH, where the two numerical models are coupled). This interface can be chosen to be at $R_{\sS\sC/\sI\sH}=0.1$ AU where the solar wind is already supersonic and super-{\alf}ic.  Below this interface the time in the boosted frame coincides with that in the rest frame:
\beq
t=t_b,\qquad \mathbf{x}=\mathbf{x}_b, \qquad R\le R_{\sS\sC/\sI\sH}
\eeq
At larger distances, $R>R_{\sS\sC/\sI\sH}$, the rest frame time exceeds the boosted frame\footnote{We note again that the analogy with the relativistic boost is far from perfect: although we took the idea of relativity of simultaneity from the first of Eqs.(\ref{eq:lorentz1}), in our case there is no relative motion of frames, since there is no time-dependent coordinate transformation, such as in the second of Eqs.(\ref{eq:lorentz1}).} time by an $R$ dependent offset: 
\beq\label{eq:tIH}
t = t_b + \Delta t(R), \qquad \Delta t(R)=\Delta t_0 \frac{R-R_{\sS\sC/\sI\sH}}{1{\rm AU}-R_{\sS\sC/\sI\sH}}, \qquad \mathbf{x}=\mathbf{x}_b,
\eeq
where $\Delta t_0=\Delta t(R=1{\rm AU})$ is the desired forecast interval at 1 AU. While the simulated time in the SC domain, $t=t_b$, is approximately an hour behind the physical time, the time at 1 AU, $t(R=1{\rm AU})=t_b + \Delta t_0$ is ahead of the physical time. 

\begin{figure}[ht]
\centering
\includegraphics[width=1\linewidth]{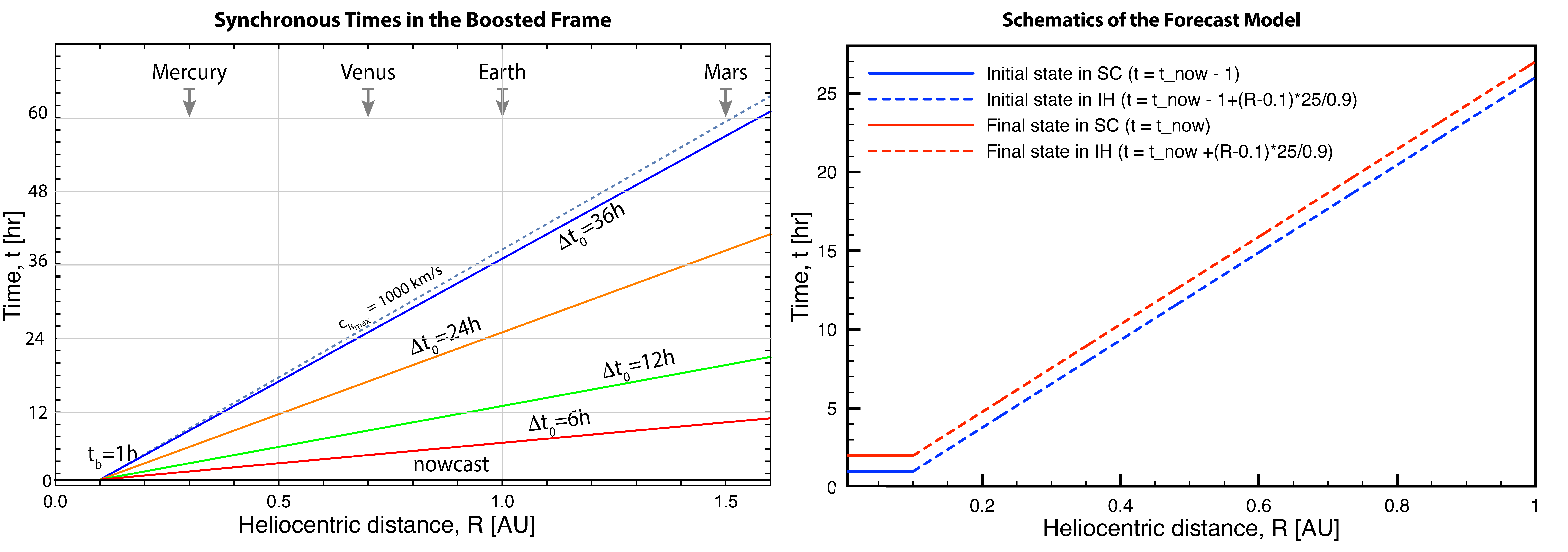}
\caption{
\textbf{Left panel:} Illustration of simultaneity in the boosted frame for different values of $\Delta t_0$. Above the horizontal (nowcast) solid black line the colored lines present predictions of future times. The black dashed line shows the limiting case where $\Lambda_m$ equals to the assumed maximum radial transient speed of $c_{R_\text{max}}=1000$ km/s, in the given state of the solar wind, thus violating inequality~(\ref{eq:speeding}). The region between the two black lines represent the domain where our proposed new method is meaningful and applicable: in this region simulations can predict ``future'' solutions \textit{not based} on future observations or assumptions about them.
\textbf{Right panel:} Space-time diagram for the proposed forecast model. The simulation time in SC ($R_\odot\le R \le R_{\sS\sC/\sI\sH}$ and $t_{\rm start}\le t_{\sS\sC}\le t_{\rm final}$) corresponds to the physical time, while in the IH model ($R_{\sS\sC/\sI\sH}\le R \le 1{\rm AU}$) the boosted frame ($t_{\rm start}+\Delta t(R) \le t_{\sI\sH}\le t_{\rm final}+\Delta t(R)$) is used. Here we used ($t_{\rm start}=t_{\rm now}-1{\rm h}$ and $t_{\rm final}=t_{\rm now}$, where $t_{\rm now}$ is the time of the latest magnetogram.}
\label{fig:Cartoon}
\end{figure}

Fig.~\ref{fig:Cartoon} summarizes the basic ideas of the proposed physics-based forecast system. Both panels show synchronous lines of events in the boosted frame of reference, \ie the lines of equal $t_b=const$, in the $(R,t)$ plane. Since we solved the SC region in the rest frame ($t=t_b$), the synchronous event line in the $R<R_{\sS\sC/\sI\sH}=0.1$ AU region is horizontal.  In the IH region ($R>R_{\sS\sC/\sI\sH}$)  the slope of the synchronous event line is controlled by the desired forecast time at 1 AU, $\Delta t_0$. The left panel demonstrates the impact of various values of $\Delta t_0$ on the final state after an hour of boosted frame simulation. Above the horizontal (now-cast) solid black line, the colored lines present predictions of \textit{future times}. We assumed above that the solar wind at $R=R_{\sS\sC/\sI\sH}$ is supersonic and super-\alf ic, so that the nowcast state in IH at $t=1$h (solid black line in the left panel of Fig.~\ref{fig:Cartoon}) is fully determined by the previous evolution of the boundary values at $R=R_{\sS\sC/\sI\sH},\,t<1$h, for about 5 days. Alternatively, the nowcast state at $t=1$h can be fully determined by the state at $R\ge R_{\sS\sC/\sI\sH},\,t=0$ and the 1 hour evolution of the boundary values at $R=R_{\sS\sC/\sI\sH},\,0\le t\le1$ h. However, the same statements are true for all colored synchronous event lines in the left panel of Fig.~\ref{fig:Cartoon} thus allowing data-driven forecast.

The choice of offset time is constrained since the inverse of its derivative, 
$\mathrm{d}[\Delta t(R)]/\mathrm{d}R$, 
determines the limiting 
speed, $\Lambda_m$:
\beq
\frac{{\mathrm{d}}(\Delta t)}{\mathrm{d}R} = \frac{\Delta t_0}{1\mathrm{AU}-R_{\sS\sC/\sI\sH}} = \frac{1}{\Lambda_m}.
\eeq
For $\Delta t_0=25$h we get $\Lambda_m\approx1,500$ km/s, which 
must exceed the maximum radial speed of transients, $c_{R_\text{max}}$, in the given state of the solar wind,
\beq\label{eq:speeding}
c_{R_\text{max}}<\Lambda_m,
\eeq
to be applicable to the proposed forecast model.
In fact, if a transient with radial speed, $c_{R_\text{max}}>\Lambda_m$ crosses the SC / IH boundary at the time instant $t=t_0$, it will reach 1 AU at $t_0 + \left(1{\rm AU} - R_{\sS\sC/\sI\sH}\right) /c_{R_\text{max}} < t_0+\left(1{\rm AU}-R_{\sS\sC/\sI\sH}\right)/\Lambda_m=t_0 + \Delta t_0$. This time is \textit{in the past} for which a forecast is already available, violating causality. Since the speed of the fastest CMEs exceeds $1,500$ km/s, one might use a higher limiting speed to avoid acausality, such as $\Lambda_m=3,000$ km/sec. However, such a large value of $\Lambda_m$ would reduce the applicable forecast time by a factor of 1/2. Inequality~(\ref{eq:speeding}), which expresses the principle of causality, is one of the criteria for the applicability of the proposed governing equations and associated numerical methods.

 The black dashed line shows the synchronous event line where $\Lambda_m = c_{R_\text{max}}$, assuming a maximum radial perturbation speed of $c_{R_\text{max}}=1000$ km/s. At and above this line inequality~(\ref{eq:speeding}) is not satisfied, and the proposed model is not applicable. The region between the two black lines represents the domain in which our proposed new method is meaningful and applicable. In this region, simulations can predict ``future'' solutions \textit{not based} on future observations or assumptions about them. For example, it can be seen that on Mars one can achieve a forecast time of 60 hours.

The right panel shows the space-time diagram for the proposed forecast simulation. It is assumed that new, updated magnetograms are available on an hourly basis, therefore, the start and end times of the simulation are $t_{\rm start} = t_{\rm now} - 1{\rm h}$ and $t_{\rm final}=t_{\rm now}$, $t_{\rm now}$ being the time of the latest magnetogram. In the boosted frame at 1 AU the initial time is $t_b(1AU) = t_{\rm now} + \Delta t_0 - 1\rm{h} = t_{\rm now} + 24\rm{h}$ (assuming a one-day forecast at 1 AU starting from the previous magnetogram, $\Delta t_0=25$h). We note that every point along the boosted frame synchronous line corresponds to a different physical time: at 1 AU it gives a $\Delta t_0-1\rm{h}=24\rm{h}$ forecast. 

Our proposed approach makes it possible to simulate the ``future'' solar wind. In the rest frame, we need to simulate about three days of physical time to properly propagate the conditions represented by a synchronic magnetogram to Earth orbit. In the boosted frame, however, the future impact of a given magnetogram on the solar wind at 1AU can be obtained about a day ahead of the physical time the solar wind parcel reaches Earth. In this way, our proposed method results in a significant actual forecast capability.

It is relatively straightforward to apply/implement the proposed framework to a system of \textit{conservation laws}, which is actually a set of partial differential equations (PDEs) of a special kind, mathematically expressing the conservation of physical quantities such as mass, momentum, and energy. Specifically, for each of these conserved  quantities the \textit{conservation laws} can be written as
\beq\label{eq_conservationlaw1}
\frac{\partial U}{\partial t}+\divg\mathbf{F}=0
\eeq
where $U$ is the density of a \textit{conserved variable} and $\divg\mathbf{F}$ is the divergence of the flux function $\mathbf{F}$ (at a given time). 

In the computational domain of SC, the equations will be solved in a frame boosted with $\Delta t_0=0$ (rest frame) and can be rewritten in terms of $t_b=t$:
\beq\label{eq:conservationlaw1a}
\frac{\partial U}{\partial t_b}+\divg\mathbf{F}=0,\qquad R_\odot\le R\le R_{\sS\sC/\sI\sH}.
\eeq
In the IH domain it is convenient to apply the coordinate transformation $\left(t,\mathbf{x}\right)\to\left(t_b,\mathbf{x}\right)$ for Eq.~(\ref{eq_conservationlaw1}). The spatial derivatives at constant $t$ are converted to those at constant $t_b$ the following way:
\begin{eqnarray}
\label{eq:coordtrans}
\renewcommand{\arraystretch}{2}
&&\frac\partial{\partial \mathbf{x}}\left(\dots\right)_{t=const} =
\frac\partial{\partial \mathbf{x}}\left(\dots\right)_{t_b=const}
+\left(\frac{\partial t_b}{\partial \mathbf{x}}\right)_{t=const}\frac\partial{\partial t_b}\left(\dots\right)_{\mathbf{x}=const}, \nonumber\\
&&\left(\frac{\partial t_b}{\partial \mathbf{x}}\right)_{t=const} = \left(\frac{\partial \left(t-\Delta t\right)}{\partial \mathbf{x}}\right)_{t=const}=-\grad(\Delta t)=-\frac\er{\Lambda_m} 
\renewcommand{\arraystretch}{1.}
\end{eqnarray}
$\er$ being a unit vector in the radial direction. With the use of Eq.~(\ref{eq:coordtrans}) the governing equations in IH can be written as follows:
\beq\label{eq_conservationlaw2}
\frac{\partial}{\partial t_b}\left(U-\frac1{\Lambda_m}\er\cdot\mathbf{F}\right) + \divg\mathbf{F}=0,\qquad R_{\sS\sC/\sI\sH}\le R\le 1{\rm AU}.
\eeq
We note that while the state variable is modified in Eq.~(\ref{eq_conservationlaw2}), the flux function remains the same. This fact will greatly simplify the resulting prediction algorithm.

\section{Characteristic Properties of the Modified Governing Equations}
\label{Sec:characteristic}

Although the system of conservation laws given by Eq.~(\ref{eq_conservationlaw1}) is non-linear in the important cases of hydrodynamics \citep{Landau1959} or magnetohydrodynamics (MHD) \citep{Shore1992}, its properties are essentially characterized by the linearized \textit{characteristic equations} for small perturbations. 

For linear waves propagating in an arbitrary direction (not necessarily radially), we can use their property of \textit{phase invariance}, well known for \textit{harmonic waves}. Consider a perturbation of primitive variables, $\delta\bm{{\cal P}}$ in the rest frame which depends on coordinates and time via the harmonic factor, $\delta\bm{{\cal P}}\propto\Re\left\{\exp\left[\mathrm{i}\left(\mathbf{k}\cdot\mathbf{x}-\omega t\right)\right]\right\}$, $\mathrm{i}$ being the imaginary unit, $\mathbf{k}$ being the wave vector and $\omega$ being its frequency. 

The vector of characteristic perturbation  velocities, $\bm{\lambda}^{(\ell)}$,  can be expressed by wave vector and frequency: $\omega=\bm{\lambda}^{(\ell)}\cdot\mathbf{k}$, $\bm{\lambda}^{(\ell)}= \partial\omega/\partial\mathbf{k}$. Phase invariance \cite[see, \eg ][]{landau1975} means that in the boosted frame (Lorenz transformation), the wave phase does not change once expressed in terms of the wave vector $\mathbf{k}_b$ and frequency $\omega_b$ in the transformed space-time: $\mathbf{k}\cdot\mathbf{x}-\omega t=\mathbf{k}_b\cdot\mathbf{x}_b-\omega_b t_b$. From this equation, with the help of Eq.~(\ref{eq:tIH}), we can find the wave vector and frequency in the boosted frame:
\beq\label{eq:3Dspeed}
\omega_b = \omega,\qquad \mathbf{k}_b = \mathbf{k}-\frac{\omega}{\Lambda_m}\er, \qquad \mathbf{k} = \mathbf{k}_b+\frac{\omega}{\Lambda_m}\er, 
\eeq
while the perturbation velocity in the boosted frame can be found from the following identity:
\beq
\bm{\lambda}^{(\ell)}_b=\frac{\partial\omega_b}{\partial\mathbf{k}_b}=\frac{\partial\omega}{\partial\mathbf{k}_b}=\frac{\partial\omega}{\partial\mathbf{k}}\cdot\frac{\partial\mathbf{k}}{\partial\mathbf{k}_b}=\bm{\lambda}^{(\ell)}\cdot
\frac{\partial\left(\mathbf{k}_b+\er\frac{\omega_b}{\Lambda_m}\right)}{\partial\mathbf{k}_b}=\bm{\lambda}^{(\ell)}+\frac{\er\cdot\bm{\lambda}^{(\ell)}}{\Lambda_m}\bm{\lambda}^{(\ell)}_b,
\eeq
giving:
\beq\label{eq:boostedspeed}
\bm{\lambda}^{(\ell)}_b=\frac{\bm{\lambda}^{(\ell)}}{1-\frac{\er\cdot\bm{\lambda}^{(\ell)}}{\Lambda_m}}
\eeq

For waves propagating outward from the Sun, \ie for $\er\cdot\bm{\lambda}^{(\ell)}>0$, the waves propagate faster in the boosted frame than in real time, thus providing the mathematical foundation for forecasting their arrival to 1 AU. This point can be illustrated with a special case of ``characteristic perturbations'', namely, the contact discontinuity bounding a ``magnetic cloud,'' which is believed to be one of the possible kinds of CMEs. The travel time in the rest frame for the CME from $R_{\sS\sC/\sI\sH}$ to a given heliocentric distance, $R$, can be expressed in terms of the propagation speed in the radial direction, $\lambda_R$: $t_{\rm travel}=\frac{R-R_{\sS\sC/\sI\sH}}{\lambda_R}$.
In the forecast model (boosted frame) the transient propagates faster (in accordance with Eq.~\ref{eq:boostedspeed}), and  consequently, its travel time to 1 AU is shorter by the chosen offset time, $\Delta t(R)$:
\beq\label{eq:traveltime}
\left(t_{\rm travel}\right)_b =\frac{R -R_{\sS\sC/\sI\sH}}{\lambda_R/(1-\lambda_R/\Lambda_m)} =\frac{R-R_{\sS\sC/\sI\sH}}{\lambda_R}\left(1 -\lambda_R/\Lambda_m\right)=t_{\rm travel}-\Delta t(R)
\eeq
Hence, once the CME is simulated in the boosted frame, its arrival time forecast can be obtained by the offset time, $\Delta t(R)$, ahead of real time. 

\begin{figure}[ht!]
\includegraphics[width=1\linewidth]{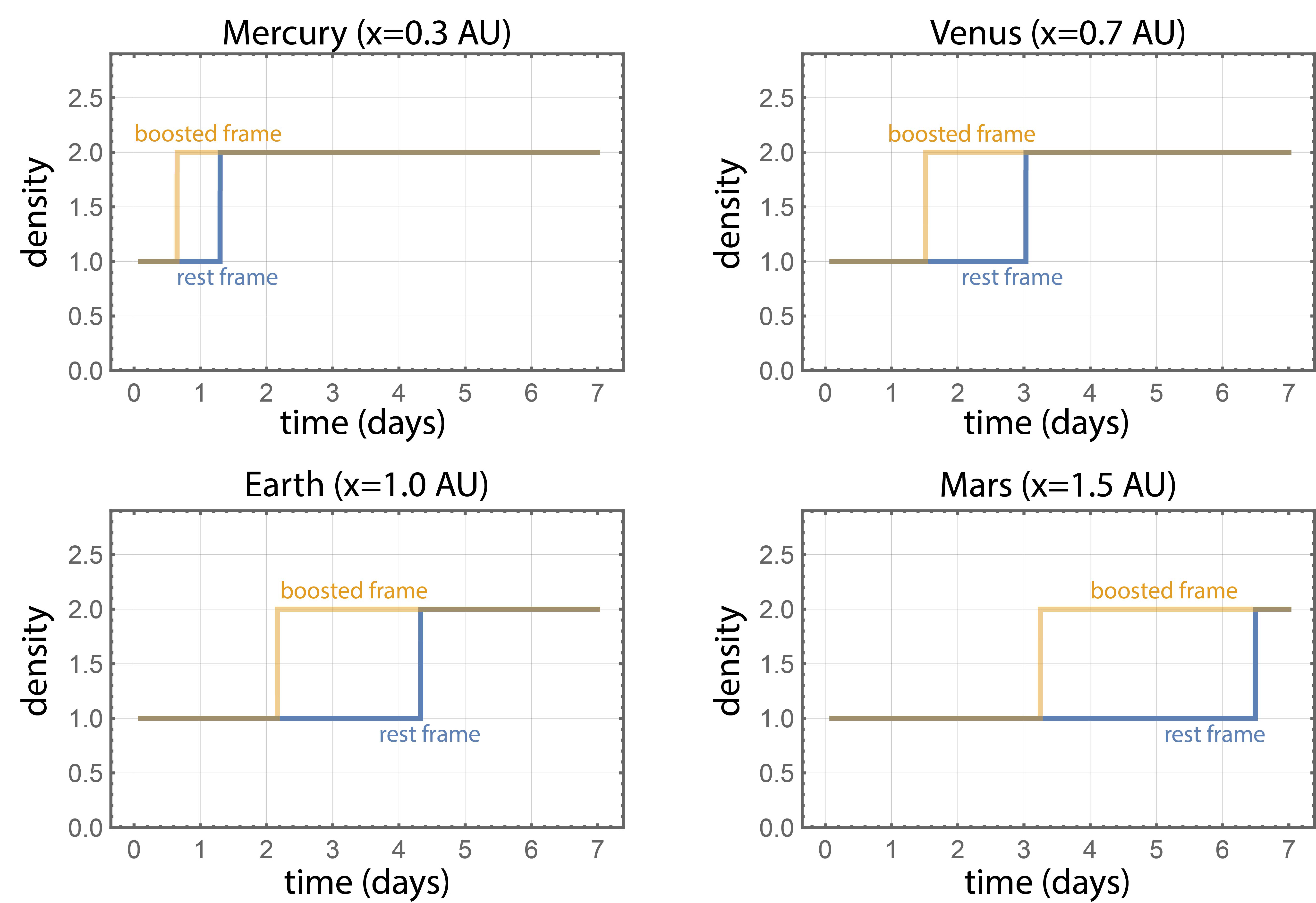}
\caption{Time evolution of the density at several radial distances following a sudden density increase at $x=0.1$ AU at $t=0$.}
\label{fig:future}
\end{figure}
As a proof of concept, let us consider a specific example of a density jump in the solar wind propagating with the speed of $\lambda_R = 400$ km/s (typical slow solar wind speed) in the rest frame and in the boosted frame with the choice of $\Lambda_m =800$ km/s (such that the propagation speed in the boosted frame is $(\lambda_R)_b=\lambda_R/(1-\lambda_R/\Lambda_m)=800$ km/s).   
Fig.~\ref{fig:future} shows the results in the orbits of the terrestrial planets. One can see that the forecast time at Mercury is less than a day, at Venus it is a day and a half, while at Earth the forecast time is about two days. This forecast time inversely depends on the limiting speed $\Lambda_m$. For the solar wind with larger maximum transient radial speed, $c_{R_\mathrm{max}}$, the limiting speed, $\Lambda_m$, will also be larger, and accordingly, a shorter forecast time should be applied (see Eq.~\ref{eq:speeding} and Fig.~\ref{fig:Cartoon} above). 
\section{Forecast System Based on the MHD Equations}
\label{Sec:mhd}
To model the solar-terrestrial environment that is significantly affected by the solar, interplanetary, or planetary magnetic field, $\mathbf{B}$, a magnetohydrodynamics (MHD) model is used. While solving MHD equations, the system of conservation laws should be completed with the source terms proportional to $\divg\bB$ \cite[see][]{powell1999}: 
\beq\label{eq:mhd0}
\frac\partial{\partial t}\left[\begin{array}{c}\rho\\\rho\bu\\ \bB\\ \frac{\rho u^2}2+\frac{P}{\gamma-1}+\frac{B^2}{2\mu_0}\end{array}\right]
+\divg\left[\begin{array}{c}\rho\bu\\\rho\bu\bu +\left(P+\frac{B^2}{2\mu_0}\right)\mathbf{I} -\frac{\bB\bB}{\mu_0}\\ \bu\bB-\bB\bu \\ \bu\left(\frac{\rho u^2}2+\frac{\gamma P}{\gamma-1}+\frac{B^2}{\mu_0}\right)-\bB\frac{\bu\cdot\bB}{\mu_0}\end{array}\right]+\left[\begin{array}{c}0\\ \frac{\bB}{\mu_0} \\ \bu \\ \frac{\bu\cdot\bB}{\mu_0}   \end{array}\right]\divg\bB=0.
\eeq
This approach allows for constructing a high-resolution numerical flux, including a non-degenerate ``8th wave'' flushing away nonzero $\divg\bB$, if any. In the next step we transform Eqs.~(\ref{eq:mhd0}) to a boosted frame using Eq.~(\ref{eq:coordtrans}). In addition to the transformation of the flux divergence described above where we derived Eq.~(\ref{eq_conservationlaw2}), the gradient operator at constant $t$ in the $\divg\bB$ term should also be expressed in spacetime $(t_b,\mathbf{x})$:
\beq\label{eq:divB}
\left(\divg\bB\right)_{t=\mathrm{const}} =\left(\divg\bB\right)_{t_b=\mathrm{const}}-\frac{1}{\Lambda_m} \frac{\partial \er\cdot\bB}{\partial t_b}.
\eeq
With these regards, in the boosted frame the revised Eq.~(\ref{eq:mhd0}) differs from Eq.~(\ref{eq_conservationlaw2}) by the terms resulting from
Eq.~(\ref{eq:divB}):
\beq\label{eq:mhd1}
\frac{\partial}{\partial t_b}\left(U_i-\frac1{\Lambda_m}\er\cdot\mathbf{F}_i\right)-\frac{1}{\Lambda_m}\frac{\partial \er\cdot\bB}{\partial t_b}\left[\begin{array}{c}0\\ \frac{\bB}{\mu_0} \\ \bu \\ \frac{\bu\cdot\bB}{\mu_0}   \end{array}\right]  + \divg\mathbf{F}_i+\left[\begin{array}{c}0\\ \frac{\bB}{\mu_0} \\ \bu \\ \frac{\bu\cdot\bB}{\mu_0}   \end{array}\right]\divg\bB =0,
\eeq
where, for brevity, we use index notations, $U_i$ and $\mathbf{F}_i$, for the conserved variables and their fluxes as explicitly defined in Eq.~(\ref{eq:mhd0}).

\section{Numerical Example for Solar Wind Forecast}
\label{Sec:mhd-example}
As a proof of principle, we simulated the CME propagation in the rest frame and boosted frame. We used the settings recently used by \cite{Liu2024} to model the April 11, 2013 CME event. We simulate SC with the \alf~wave turbulence-driven solar atmosphere model in real time (AWSoM-R) as described in \cite{Sokolov2021}. However, for the IH domain we used the simplified MHD model of Sect.~\ref{Sec:mhd} accomplished with adiabatic equation. 
for electron pressure, with no turbulence.

\begin{figure}[ht]
\centering
\includegraphics[width=1\linewidth]{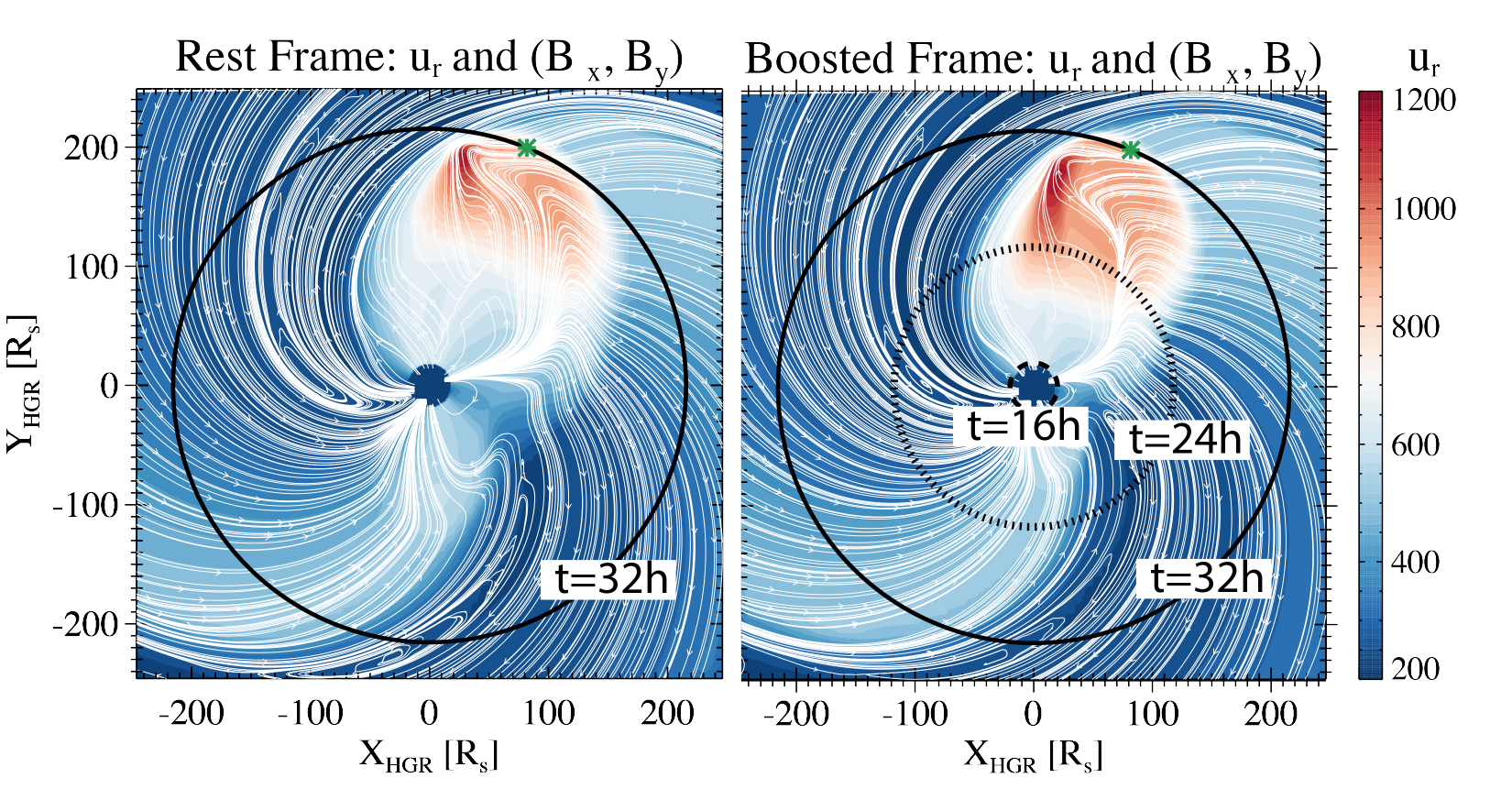}
\caption{Left panel: Snapshot of the time accurate solution in the rest frame 32 hours after the CME onset; Right panel: Boosted frame solution at $t_b = t-\frac{16}{0.9} (R-0.1)$ ([t]=h, [R]=AU). The plasma and IMF parameters at the concentric circles represent solutions at t= 16h, 24h and 32h after CME onset. Earth's location is marked by a green asterisk. The two solutions are identical at 1AU (the $t=32$h circle), but differ at all other distances. Note that the boosted solution at 1AU is 16h ahead of the solution at the SC/IH boundary.}
\label{fig:boost-compare}
\end{figure}

The ambient (background) state in the coupled SC and IH models is obtained as a steady-state solution of the stream-aligned MHD equations by \cite{Sokolov2022stream}. After obtaining the steady-state solution, we applied the Eruptive Event Generator using Gibson-Low configuration \cite[EEGGL, see][and references therein]{borovikov17, Jin:2017a}. We simulated CME propagation in the rest frame for $t=32$ hours of physical time after the CME onset. The resulting state in the IH for the final time is presented in the left panel of Fig.~\ref{fig:boost-compare}. The white lines are the magnetic field lines and the color bar shows the distribution of radial solar wind speed in the solar equatorial plane, \ie in the $xy-$coordinate plane of the HGR coordinate system. It can be seen that at $t\approx32$ hours the CME-driven shock reaches the Earth location (shown by the green asterisk).

In the right panel, we provide the result of a simulation, in which the SC is modeled in the rest frame, however, in the IH the boosted frame is used. The panel shows the $xy-$plane in HGR coordinates at time $t_b=16$ hours after CME onset. 

With the choice of the time offset of $\Delta t_0=25$ hours at 1 AU, the scheme described in the appendices is applicable for most of the simulation. However, in the time interval of $t_b\approx2-4$ hours, the plasma radial speed exceeds $1600\mathrm{km/s} >\Lambda_m\approx1500\mathrm{km/s}$. The optimal choice of time offset to simulate this event in the boosted frame appears to be $\Delta t_0=16$ hours. In this case, the maximum boost factor is as high as $\Gamma_B\ge10$ at $t_b\approx2$, which means that the maximum radial speed of the perturbations, $c_{R_\mathrm{max}}$, is very close to $\Lambda_m$. With this choice of time offset, the physical time and the boosted time are identical at $t=t_b=16$ hours at a heliocentric distance of $0.1$ AU (shown by the dashed circle). At the heliospheric distance of 0.55 AU (shown by a dotted line), half of the offset time is added, hence the data correspond to the time of $t=t_b+\Delta t_0/2=24$ hours after the CME onset. At 1 AU, shown by the solid circle, the simulated parameters correspond to the physical time with the full time offset, $t=t_b+\Delta t_0=32$ hours, and the plot clearly demonstrates the arrival of the shock wave to the Earth location (green star) at $t=32\mathrm{km/s}$. 

Thus, in the full numerical test, presented in Fig.~\ref{fig:boost-compare}, the shock wave arrival time is obtained in two ways: in the traditional (rest) frame and in the newly proposed (boosted) frame. Both approaches provide identical predictions for results at 1 AU, thus justifying both the quantitative accuracy of the new method and the reliability of the numerical algorithms described in the appendices. 

\section{Summary}
Inspired by the concept of \textit{relativity of simultaneity} used in the theory of special relativity, a new model is developed to simulate \textit{future} solar wind conditions at any point in the inner solar system. It is based on time transformation between two coordinate systems: the solar rest frame and a boosted system in which the current state of the solar wind at the solar corona -- inner heliosphere boundary (where the solar wind is supersonic and super-{\alf}ic) and future states of the solar wind in the IH are simultaneous. We derived the modified governing equations for both hydrodynamics (HD) and magnetohydrodynamics (MHD) and presented a new numerical algorithm that solves the modified governing equations. The proposed method enables an efficient numerical implementation and thus a significantly longer forecast time than traditional solution methods.

An important distinctive feature of the proposed approach is that the simulation in the SC is driven by hourly updated solar magnetograms and is continuously simulated in nearly real time. The predictive capability for the solar wind and IMF parameters at 1 AU is achieved by using the time offset, such that current observations of the solar magnetic field are simulated simultaneously with tomorrow's parameters of the ambient solar wind at 1 AU.  In the numerical test for transient propagation, the boosted solution for the CME-driven shock arrival at 1AU is 16 hours ahead of the solution at the SC/IH boundary.

\section{Acknowledgements}
The authors thank Maria M. Kuznetsova (NASA GSFC) for stimulating our interest in the space weather forecast problem and acknowledge discussions with Gabor Toth and Lulu Zhao as well as the help of Weihao Liu with visualization. This research is partially based on work supported by a NASA LWS Strategic Capability (SCEPTER) project at the University of Michigan under NASA grant 80NSSC22K0892, by a NASA LWS Space Weather Centers of Excellence project (CLEAR) at the University of Michigan under NASA grant 80NSSC23M0191, by NASA grant 80NSSC21K1124, by NSF ANSWERS grant GEO-2149771, and by NASA LWS grant 80NSSC20K1778. We also acknowledge high-performance computing support from: (1) Yellowstone provided by NCAR's Computational and Information Systems Laboratory, sponsored by the NSF, and (2) Pleiades operated by NASA's Advanced Supercomputing Division. 

\bibliographystyle{aasjournal}

\appendix
\section{One-dimensional Problems in Boosted and Rest Frames }
\label{Sec:1Dproblem}

For waves propagating in the radial direction, the characteristic form of Eqs.~(\ref{eq_conservationlaw1}) reads
\beq
\frac{\partial U_i}{\partial t}+\sum_j{\frac{\partial\left(\er\cdot\mathbf{F_i}\right)}{\partial U_j}\cdot\frac{\partial U_j}{\partial R}}=0,
\eeq
where the conserved variables and their flux vectors are denoted with the subscript index $i$.
If the system of conservation laws is \textit{hyperbolic}, the Jacobian matrix, $\partial\left(\er\cdot\mathbf{F_i}\right)/\partial U_j$, possesses a full set of eigenvalues, $\lambda^{(\ell)}$, and corresponding eigenvectors. Each of the eigenvectors describes some linear combinations of (small increments in) conserved variables, which are referred to as {\it the Riemann invariants}, ${\mathcal{R}}^{(\ell)}$ and obey a set of independent linear advection equations:
\beq\label{eq:chareq}
\frac{\partial {\mathcal{R}}^{(\ell)}}{\partial t}+\lambda^{(\ell)}\frac{\partial {\mathcal{R}}^{(\ell)}}{\partial R}=0,
\eeq
in which the Jacobian eigenvalue plays the role of the \textit{characteristic speed}. The solution of this kind of equation is any function ${\mathcal R}(R/\lambda^{(\ell)}-t)$ of a single argument $(R/\lambda^{(\ell)}-t)$, which describes the wave propagation along the spacetime line, obeying the equation:
\beq\label{eq:charline1}
\frac{\mathrm{d}t}1=\frac{\mathrm{d}R}{\lambda^{(\ell)}}.
\eeq
The denominators in Eq.~(\ref{eq:charline1}) are equal to the factors multiplying the time and coordinate derivatives in Eq.~(\ref{eq:chareq}). 

Although the above considerations are directly applicable to the SC domain, for the modified system of conservation laws used in the IH domain, the characteristic equation (Eq.~(\ref{eq_conservationlaw2}) needs to be modified:
\beq\label{eq:chareq1}
\left(1-\frac{\lambda^{(\ell)}}{\Lambda_m}\right)\frac{\partial {\mathcal{R}}^{(\ell)}}{\partial t_b}+\lambda^{(\ell)}\frac{\partial {\mathcal{R}}^{(\ell)}}{\partial R}=0,
\eeq
which results in the modified equation for the characteristic spacetime line
\beq\label{eq:charline2}
\frac{\mathrm{d}t_b}{1-\frac{\lambda^{(\ell)}}{\Lambda_m}}=\frac{\mathrm{d}R}{\lambda^{(\ell)}},
\eeq
corresponding to the modified characteristic speed of $\lambda^{(\ell)}
/(1-\lambda^{(\ell)}/\Lambda_m)$. Eq.~(\ref{eq:charline2}) yet again demonstrates the need for the limiting speed, $\Lambda_m$ given by Eq.~(\ref{eq:speeding}), since if any eigenvalue, $\lambda^{(\ell)}$, exceed $\Lambda_m$, the corresponding wave would propagate back in time (see Eq.~\ref{eq:charline2}). 

The analytical solution of the one-dimensional (1D) Riemann problem for the hydrodynamic equations with discontinuous initial conditions is the flow with constant parameters for $0\le x<0.5$, denoted below as the left state (subscript ``L'') and the flow with a different set of constant parameters at $0.5<x\le1$ denoted with subscript ``R.'' The time offset is zero at $0.5\le x\le 1$ and increases linearly with $x$ for $x\ge0.5$, reaching the value of $\Delta\tau_0$ at the right boundary. The evolution during $0\le t\le \Delta\tau_0$ for $0\le x\le 0.5$ is governed by the usual hydrodynamic equations,  while for $0.5\le x\le 1$ the modified governing equations, given by Eqs.~(\ref{eq_conservationlaw2}), are used with $\Lambda_m=0.5/\Delta \tau_0$. 

In hydrodynamics, the solution of the Riemann problem for the set of \textit{primitive variables}, $\bm{\mathcal{P}} =\left[\rho,\bu,p\right]^T$, that combines the density, velocity vector, and pressure, is known to be a \textit{self-similar} function $\bm{\mathcal{P}}_{\rm RS}$,  combining time and location, $(x-0.5)/t$: $\bm{\mathcal{P}} =\bm{\mathcal{P}}_{\rm RS}([x-0.5]/t)$. In the linear approximation, the solution is composed of constant states separated by jump-like waves (the \textit{simple} Riemann waves) propagating with the characteristic speeds from the initial location of the discontinuity at $x=0.5$, \ie the range of the self-similar argument for each constant state is $\lambda^{(\ell-1)}< (x-0.5)/t<\lambda^{(\ell)}$. This solution of the Riemann problem can be applied as is for the region $0\le x\le 0.5$. However, for $0.5\le x\le 1$ the speeds of characteristic waves must be modified in accordance with Eq.~(\ref{eq:charline2}), so that the self-similar argument of the $\bm{\mathcal{P}}_{\rm RS}$ function needs to be modified accordingly:
\bea\label{eq:Riemannproblem}
\renewcommand{\arraystretch}{1.75}
\bm{\mathcal{P}}=\left\{
\begin{array}{ll}
$$\bm{\mathcal{P}}_{\rm RS}\left(\frac{x-0.5}{t_b}\right)$$ & $$\text{if}\ 0\le x\le 0.5$$ \\
$$\bm{\mathcal{P}}_{\rm RS}\left(\frac{\frac{x-0.5}{t_b}}{1+\frac1{\Lambda_m}\frac{x-0.5}{t_b}}\right)$$ & $$\text{if}\ 0.5\le x\le 1.$$
\end{array}\right.
\renewcommand{\arraystretch}{1.}
\eea

This solution can also be applied to the solution of a nonlinear Riemann problem with no loss of generality because it can also be obtained from the known with a self-similar argument by applying spacetime transformation to the argument.

\begin{figure}[t]
\centering
\includegraphics[width=0.495\linewidth]{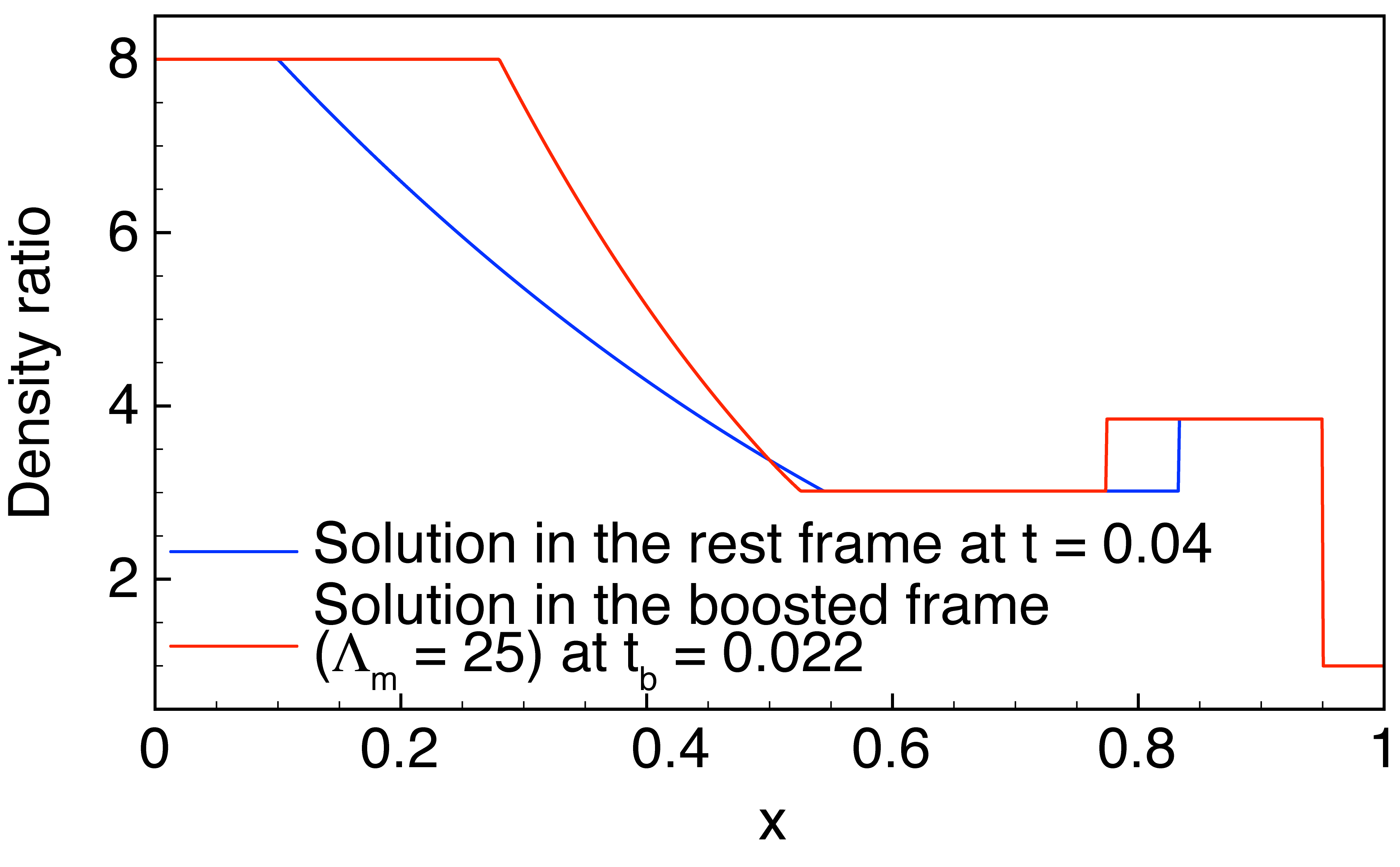}
\includegraphics[width=0.495\linewidth]{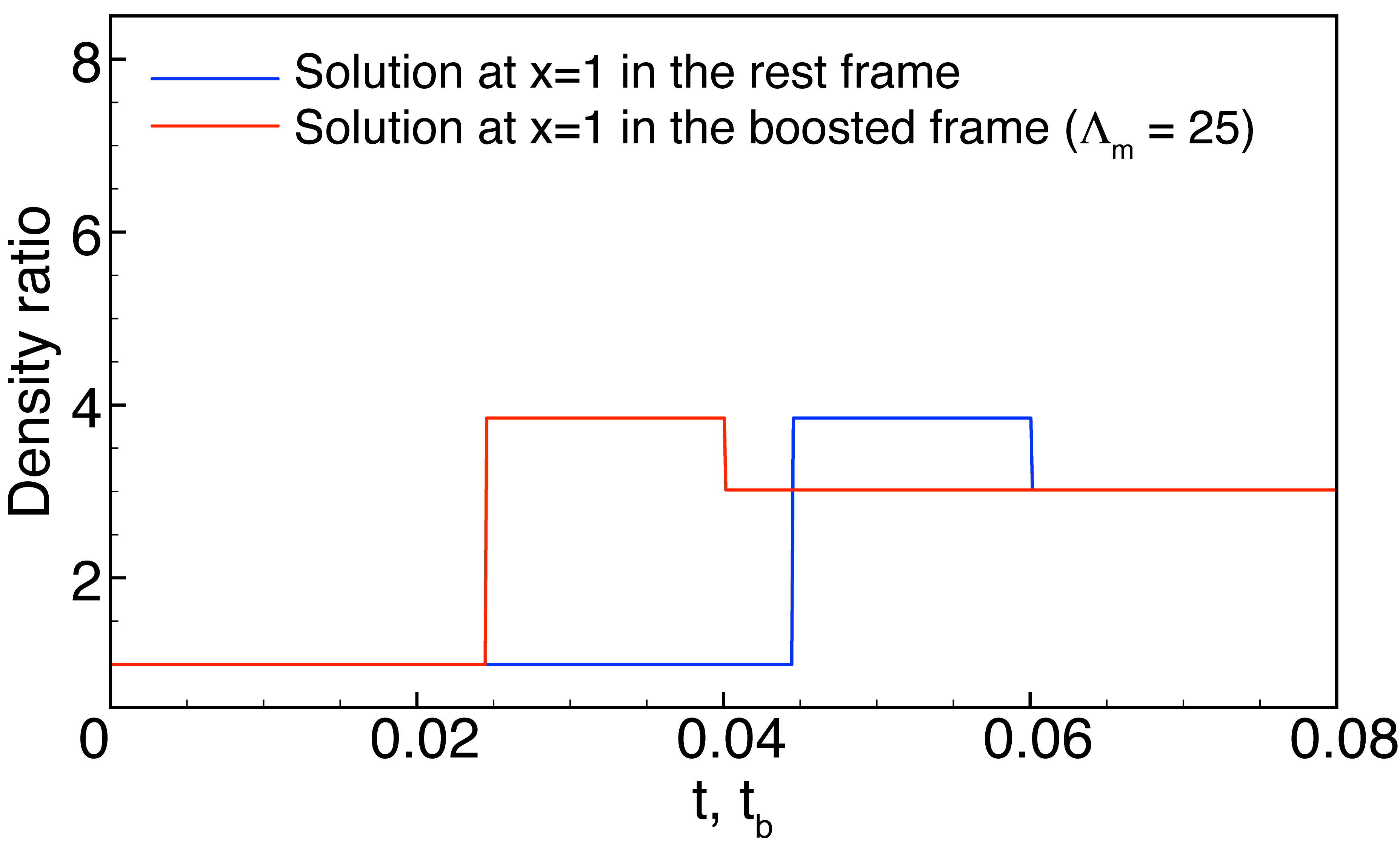}
\caption{Solution of the Riemann problem with the initial condition: $\rho=8$, $u=0$ and $p=480$ at $x<0.5$; $\rho=1$, $u=0$ and $p=1$ at $x>0.5$, for a gas with $\gamma=5/3$. {\bf Left panel:} Blue line shows the solution of hydrodynamics equations in the rest frame at the time, $t=0.04$; red line presents the solution in the boosted frame at time $t_b=0.022$. {\bf Right panel:} Time profile of density at $x=1$. Red line is the solution in the boosted frame, blue line is the solution in the rest frame. $\rho(t_b)$ predicts the solution in the rest frame, $\rho(t)$ with the time offset of $0.02$.}
\label{fig:scheme}
\end{figure}

As an example, Fig.~\ref{fig:scheme} presents the solution of the Riemann problem for a gas with a polytropic index, $\gamma=5/3$ with the following initial condition: $\rho=8$, $u=0$ and $p=480$ in $0\le x<0.5$; $\rho=1$, $u=0$ and $p=1$ for $0.5<x\le 1$.  In the left panel, the blue line shows the solution of hydrodynamic equations in the rest frame at time $t=0.04$. At this time, the rarefaction wave propagating to the left reaches $x=0.1$, while the shock wave propagating to the right arrives at the point of ``observation'' at $x=0.95$ (note that the red and blue lines overlap beyond $x=0.85$). The red line is the solution in the boosted frame, such that the time offset $\Delta t(x)$ varies from zero at $x\le 0.5$ to $\Delta t_0=0.02$ at $x=1$ (which corresponds to $\Lambda_m=(1-0.5)/0.02=25$) for the time instant, $t_b=0.022$. At $x\le0.5$ (without time offset) the rarefaction wave passes a much shorter distance during the shorter time interval, $0.022<0.04$. However, the time offset at $x=0.95$  is $(0.95-0.5)/\Lambda_m=0.018$, therefore, the shock arrival time in the boosted frame plus the time offset equals the shock arrival time in the rest frame $0.04 = 0.022+0.018$. The right panel presents the time dependence of density at $x=1$ in the rest frame (blue line) and in the boosted frame (red line). The comparison of results clearly demonstrates the time offset ($=0.02$).

\section{Finite Volume Formulation of Boosted Conservation Laws}
\label{Sec:FV}

\subsection{Discretization}
\label{Subsec:discretize}
One of the most important examples of the system of conservation laws are the hydrodynamic equations. Computational Fluid Dynamics (CFD) has been developed as a powerful applied science employing a variety of numerical methods, reviewed by \cite{Hirsch97}. Among them is the widely used \textit{finite-volume approach}. Once the system of equations given by Eq.~(\ref{eq_conservationlaw1}) is integrated over a {\it control volume}, $\Delta V$,  the integral of the term, $\divg\mathbf{F}$, in each equation reduces to a surface integral of the flux function over the boundary of the control volume:
\beq\label{eq:Gauss}
\int\limits_{\Delta V}{\left(\divg\mathbf{F}\right)\mathrm{d}^3\mathbf{x}}=\int\limits_{\sigma}{\mathbf{F}\cdot\mathrm{d}\boldsymbol{\sigma}}
\eeq

If the {\it computational domain} of the conservation-law system is decomposed into a set of control volumes ({\it cells}), $\Delta V_i$, the time evolution of the conserved variable within each control volume reduces to the exchange by the {\it numerical fluxes} between each pair of neighboring cells, $i$th and $j$th:
\beq\label{eq:scheme}
\frac{\mathrm{d}}{\mathrm{d}t}\int\limits_{\Delta V_i}{U\left(t,\mathbf{x}\right)\,\mathrm{d}^3\mathbf{x}}=-\sum_j\mathbf{F}_{ij}\cdot\bm{\sigma}_{ij}
\eeq
The numerical fluxes $\mathbf{F}_{ij}\cdot\bm{\sigma}_{ij}$,  are essentially the integrals of the flux function, $\mathbf{F}$, over the interface (the shared boundary) of two neighboring cells. Gauss' theorem (Eq.~\ref{eq:Gauss}) is formulated via  the dot product of the flux function with the ``external'' unit vector of the boundary surface for $i$th cell, which is at the same time the negative of the ``external'' unit vector to the same interface for $j$th neighboring cell, so that the numerical flux from $i$th cell to $j$th cell is always equal to the negative of the flux from $j$th cell to $i$th cell, therefore, $\bm{\sigma}_{ji}=-\bm{\sigma}_{ij}$. Consequently, the time derivative of the total integral of the conserved quantity in the computational domain, given by the sum of Eqs.~(\ref{eq:scheme}) over all control volumes, reduces to mutual canceling contributions of each numerical flux to neighboring cells, resulting in automatically conserved total physical quantities such as mass, momentum, and energy, unless there is a non-vanishing flux of these quantities through the external boundary of the computational domain. Such schemes are well known as \textit{conservative numerical schemes}. In terms of the averages of conserved variables,   over the $i$th control volume at sequentially increasing time levels with a time-step of $\Delta t$,
\beq
U^n_i=\frac1{\Delta V_i}\int\limits_{\Delta V_i}{U\left(t=t^n,\mathbf{x}\right)\,\mathrm{d^3}\mathbf{x}},
\eeq
where the superscript, $n$, denotes variables related to time $t=t^n$, and the numerical scheme becomes:
\beq\label{eq:scheme1}
U^{n+1}_i=U^n_i-\frac{\Delta t}{\Delta V_i}\sum_j\mathbf{F}_{ij}\cdot\bm{\sigma}_{ij}
\eeq
\begin{figure}[ht]
\centering
\includegraphics[width=0.5\linewidth]{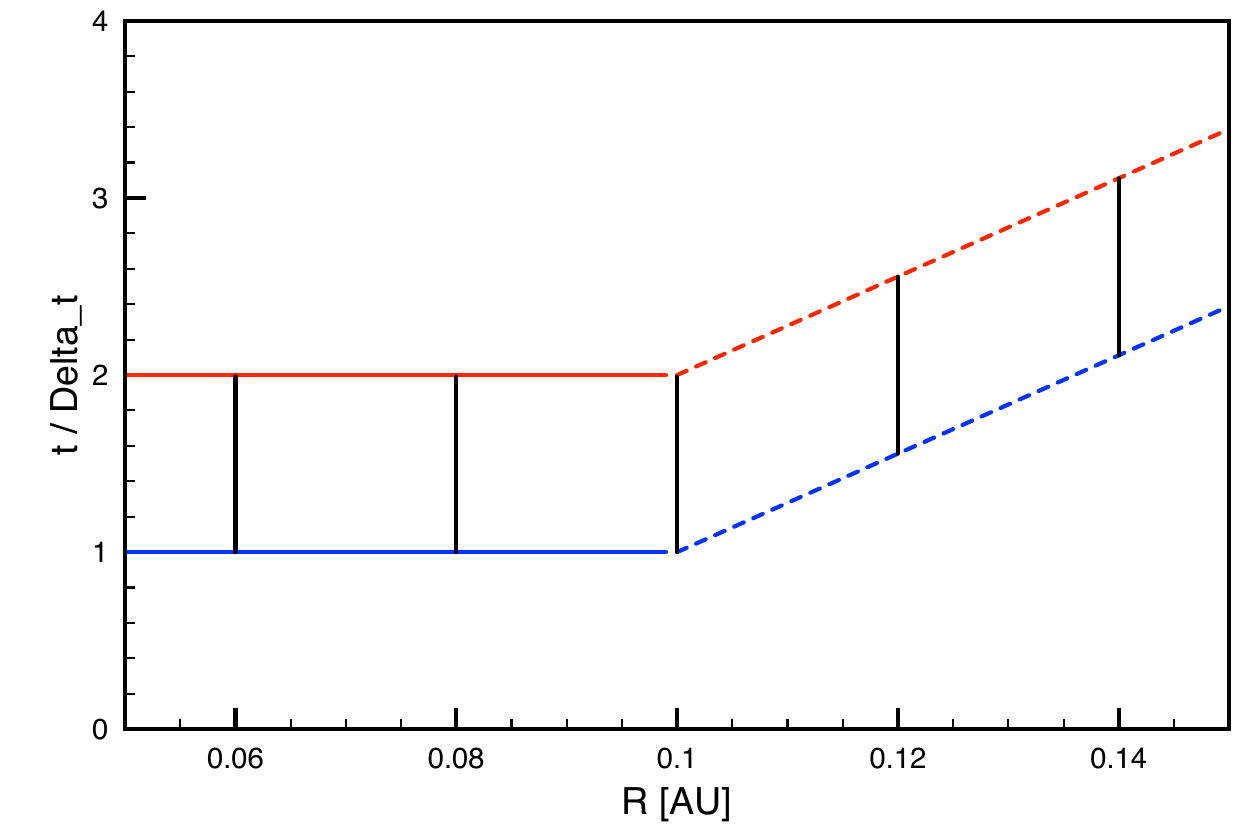}
\caption{4D control volumes used in the Lorentz boosted frame computations on the $R-t$ diagram. Vertical lines show the spatial immovable boundaries of the control volumes.  In the SC domain, $R<0.1$ AU, the time in the boosted frame coincides with that of in the rest frame and at the time levels of constant $t_b$, $t_b=t^n$ (horizontal solid blue line) and $t_b=t^{n+1}$ (horizontal solid red line) the real time is also constant, $t=t^n$ and $t=t^{n+1}$. In the IH model, at time levels $t_b=t^n$ (blue dashed line) and $t_b=t^{n+1}$ (red dashed line), the real time is not constant and the dashed lines are not parallel to the horizontal direction.}
\label{fig:volume}
\end{figure}
In applying this approach to Eq.~(\ref{eq:conservationlaw1a}) in the SC part of computational domain, we directly solve for $U^{n+1}_i$, 
\beq\label{eq:Un+1}
U^{n+1}_i =U^n_i-\frac{\Delta t_b}{\Delta V_i} \sum_j\mathbf{F}_{ij} \cdot \bm{\sigma}_{ij},\qquad R<R_{\sS\sC/\sI\sH}.
\eeq
However, in the boosted IH domain to solve Eq.~(\ref{eq_conservationlaw2}), we need to choose a special form of the 4D control volume (see Fig.~\ref{fig:volume}), in which the $n$th state at $t_b=t^n$ in the boosted frame corresponds to the \textit{hypersurface} $t-(\|\mathbf{x}\| - R_{\sS\sC/\sI\sH})/\Lambda_m=t^n$ in the rest frame 4D, $\left(t,\mathbf{x}\right)$ (blue dashed line in Fig.~\ref{fig:volume}). Integrating Eq.~(\ref{eq_conservationlaw2}) over control volume and over $t_b$ gives the finite volume formulation as follows:
\beq\label{eq:viatildeU}
\tilde{U}^{n+1}_i =\tilde{U}^n_i-\frac{\Delta t_b}{\Delta V_i} \sum_j\mathbf{F}_{ij}\cdot\bm{\sigma}_{ij}, \qquad R>R_{\sS\sC/\sI\sH},
\eeq
where the conserved quantity is modified as follows:
\beq\label{eq:tildeU}
\tilde{U}^n_i=\frac1{\Delta V_i}\int\limits_{\Delta V_i}{\tilde{U}\left(t_b=t^n,\mathbf{x}\right)\,\mathrm{d^3}\mathbf{x}}, \qquad \tilde{U}=U-\frac{\er\cdot\mathbf{F}}{\Lambda_m}.
\eeq
There is an alternative way to derive Eqs.~(\ref{eq:viatildeU}) and (\ref{eq:tildeU}) by applying Gauss' theorem to Eq.~(\ref{eq_conservationlaw1}), which can be considered as an equation for a divergence-free 4-vector, $\left(\frac\partial{\partial t}, \frac\partial{\partial \mathbf{x}} \right)\cdot\left(U,\mathbf{F}\right)=0$. The integral of this 4D divergence over the 4D control volume shown in Fig.~\ref{fig:volume} reduces to an integral over hypersurfaces bounding the control volume. Since the hypersurfaces, shown by blue and red dashed lines in Fig.~\ref{fig:volume} are no longer orthogonal to the $t$ axis, the integrand over these hypersurfaces is no longer equal to $U$, but to a linear combination of $U$ and $F_\sR=\er\cdot\mathbf{F}$. Similarly, if the (numerical) flux is calculated in the reference frame moving with a velocity of $\Lambda$, or if the control volume moves, the spatial boundary of the control volume is no longer parallel to the direction of the axis $t$, resulting in the modification of the flux function ($F\rightarrow F-\Lambda U$ in the 1D case \cite[\eg][]{Sokolov2002}).

Eqs.~(\ref{eq:Un+1}) and (\ref{eq:viatildeU}) constitute an almost complete solution algorithm, since the computation of the numerical flux is the most laborious and time consuming part of the numerical scheme and this part is taken without modification of the standard control volume approach. Specifically, an existing finite-volume code can be easily converted for solar wind forecasting. In addition to simplified computation, the use of identical numerical fluxes in the rest frame and in the boosted frame ensures the equivalence of a steady-state ``ambient'' solution for the solar wind in the coordinate frame corotating with the Sun regardless of the frame in which it is obtained. In fact, the numerical solution, obtained in the rest frame with the numerical scheme given by Eq.~(\ref{eq:scheme1}) describes steady-state, $U^{n+1}_i\equiv U^{n}_i$ for all conserved variables in all cells, if and only if $\frac{1}{\Delta V_i}\sum_j\mathbf{F}_{ij}\cdot\bm{\sigma}_{ij}\equiv0$. As long as the same numerical fluxes are used in the numerical scheme given by Eq.~(\ref{eq:viatildeU}) in the boosted frame, the solution is also steady state, since $\tilde{U}^{n+1}_i\equiv \tilde{U}^{n}_i$.  There are, however, two differences to be addressed when the boosted frame is used: the recovery of primitive variables from the modified conservative ones and the modification of the Courant condition on the time step caused by the modification of the characteristic wave speed.

\subsection{Recovery of Primitive Variables}
\label{Subsec:recovery}
Once the state vector is known at $t_b=t^n$ and the numerical fluxes are calculated, Eq.~(\ref{eq:viatildeU}) allows us to obtain the modified conserved variable, $\tilde{U}^{n+1}_i$. However, unlike in the case of standard conserved variables, it is not easy to recover physical quantities, $\bm{\mathcal{P}}^{n+1}_i$, from known modified conserved variables. One of the reasons is the non-linear dependence of $\tilde{U}^{n+1}$ on the radial component of the solar wind velocity, $u^{n+1}_\sR= \er\cdot\mathbf{u}^{n+1}$. In fact, since this velocity determines the radial component of the momentum density, it is a part of some component of $U^{n+1}$. At the same time, the flux projection on the radial direction, $(\er/\Lambda_m) \cdot\mathbf{F}^{n+1}$ usually includes the advection term, $(u^{n+1}_\sR/\Lambda_m) U^{n+1}$, so that $u^{n+1}_\sR$ is also part of all modified conserved variables, making it challenging to derive.

A similar problem is well-known in relativistic computational hydrodynamics \cite[see, \eg][]{Falle1996,Sokolov2001} where the relativistic $\Gamma$-dependence of both the density and the energy-momentum tensor components make the equation for velocity highly implicit. At the moment, there are two published ways to handle this issue. One of these is an iterative procedure to solve the primitive variables $\bm{\mathcal{P}}^{n+1}_i$ \cite[see, e.g.][``there is no difficulty in devising an efficient iterative procedure'', p. 589]{Falle1996}. The most generic way to iterate $\bm{\mathcal{P}}^{(n+1,h)}_i,\,h=0,1,2\dots$ is to start the iterative process with the value at the time level $t^n$: $\bm{\mathcal{P}}^{(n+1,0)}_i=\bm{\mathcal{P}}^{n}_i$. Successive iterations can be found by rewriting the difference in the modified conserved variables in Eqs.~(\ref{eq:Un+1}) in the form of elements of an infinite series:
\beq
\begin{split}
& \sum\limits_{h=0}^{\infty} {\left[\tilde{U}\left(\bm{\mathcal{P}}^{(n+1,h+1)}_i\right) -\tilde{U}\left(\bm{\mathcal{P}}^{(n+1,h)}_i\right)\right]}=-\frac{\Delta t_b}{\Delta V_i}\sum_j\mathbf{F}_{ij}\cdot\bm{\sigma}_{ij},
\\
& \quad \bm{\mathcal{P}}^{(n+1,0)}_i =\bm{\mathcal{P}}^n_i,\quad \bm{\mathcal{P}}^{(n+1,\infty)}_i =\bm{\mathcal{P}}^{n+1}_i.
\end{split}\label{eq:series}
\eeq
In each step of the iteration procedure, the equations are solved as follows:
\beq\label{eq:recover}
\tilde{U}\left(\bm{\mathcal{P}}^{(n+1,h+1)}_i\right) -\tilde{U}\left(\bm{\mathcal{P}}^{(n+1,h)}_i\right) =\Delta^{h},
\eeq
where
\beq\label{eq:endefect}
\Delta^{h}=
\renewcommand{\arraystretch}{2.}
\left\{\begin{array}{l}
-\frac{\Delta t_b}{\Delta V_i}\sum_j\mathbf{F}_{ij}\cdot\bm{\sigma}_{ij},\quad\qquad\qquad h=0\\
\Delta^{h-1}+\tilde{U}\left(\bm{\mathcal{P}}^{(n+1,h-1)}_i\right) -\tilde{U}\left(\bm{\mathcal{P}}^{(n+1,h)}_i\right),\quad h\ge1
\end{array}
\renewcommand{\arraystretch}{1.}
\right.
\eeq

The left hand side (LHS) of Eqs.~(\ref{eq:recover}) can be approximately linearized as 
\beq\label{eq:linearized}{
\tilde{U}\left(\bm{\mathcal{P}}^{(n+1,h+1)}_i\right) -\tilde{U}\left(\bm{\mathcal{P}}^{(n+1,h)}_i\right)\approx \frac{\partial\tilde{U} \left(\bm{\mathcal{P}}^{(n+1,h)}_i\right)}{\partial \bm{\mathcal{P}}^{(n+1,h)}_i} \cdot \delta^{h+1}\bm{\mathcal{P}}_i
}\eeq
where $\delta^{h+1}\bm{\mathcal{P}}_i=\bm{\mathcal{P}}^{(n+1,h+1)}_i-\bm{\mathcal{P}}^{(n+1,h)}_i$. The set of the linearized Eqs.~(\ref{eq:recover}) for each conserved variable constitutes the full system of linear equations to solve\footnote{In \ref{Subsec:primitive} we present a simple and explicit solution for $\delta^{h+1}u_\sR$ satisfying this system} for $\delta^{h+1} \bm{\mathcal{P}}_i$ and then to find the next iteration of primitive variables:
\beq
\bm{\mathcal{P}}^{(n+1,h+1)}_i=
\bm{\mathcal{P}}^{(n+1,h)}_i+\delta^{h+1}\bm{\mathcal{P}}_i.\eeq   
Note, that it may be beneficial to start the iterative procedure from different set of primitive variables, \eg such that $U\left(\bm{\mathcal{P}}^{(n+1,0)}_i\right) =U\left(\bm{\mathcal{P}}^{n}_i\right)-\frac{\Delta t_b}{\Delta V_i}\sum_j\mathbf{F}_{ij}\cdot\bm{\sigma}_{ij}$, or the intermediate state $\bar{U}^{(n+1)}$ as present in Eq.~(\ref{eq:donorcell}) below. The alternative choice results in extra terms, $\tilde{U}\left(\bm{\mathcal{P}}^{n}_i\right)-\tilde{U}\left(\bm{\mathcal{P}}^{(n+1,0)}_i\right)$, in the RHS of Eq.~(\ref{eq:series}) as well as in Eq.~(\ref{eq:endefect}) for $h=0$.

Another way to solve for the physical quantities is to seek special cases where the primitive variables can be solved from a single algebraic equation. For example, \cite{Sokolov2001} showed that a special choice of the equation of state allows recovering the primitive variables from the conservative ones in relativistic hydrodynamics by solving a quadratic equation. Similarly, for the forecast system based on the 3D hydrodynamic equations in the boosted frame, the change in pressure through the time step, $p^{n+1}-p^n$, can be solved from a quadratic equation, as we demonstrate in \ref{Sec:hydro}.

\subsection{Revised Courant Condition}
\label{Subsec:courant}

For a 1D problem, the Courant condition for stability of the numerical algorithm is straightforward. It is controlled by the maximum perturbation speed $c_\mathrm{max}$. In the rest frame, the time step, $\Delta t$, the size of the cell, $\Delta x_i$, and the maximum perturbation speed in each cell, $c_{(\mathrm{max},i)}$, should satisfy the condition as follows:
\beq
\frac{\Delta t\, c_{(\mathrm{max},i)}}{\Delta x_i}=
\mathrm{CFL}<1,
\eeq
where $\mathrm{CFL}$ is the local value of the Courant-Friedrichs-Levi (CFL) number. With the choice of global CFL number (should be less than one), the time step for stable computations in the rest frame can be calculated as follows:
\beq\label{eq:1Dtimestep}
\Delta t = \mathrm{CFL}\min_i\left(\frac{\Delta x_i}{c_{(\mathrm{max},i)}}\right).
\eeq

For 1D solution on spherical grid in the boosted frame this formula should be modified in accordance with Eq.~(\ref{eq:charline2}) for the maximum radial perturbation speed, {$c_{R_{(\mathrm{max},i})}$:
\beq\label{eq:1dboosttimestep}
\Delta t_b = \mathrm{CFL}\min_i\left[\left(\frac{\Delta x_i}{c_{R_{(\mathrm{max},i)}}}\right)\left(1-\frac{c_{R_{(\mathrm{max},i)}}}{\Lambda_m}\right)\right].
\eeq
This time step is positive as long as Ineq.~(\ref{eq:speeding}) holds. Compared with the Courant condition in the rest frame, the smallest allowed time step in each cell should be reduced by a factor of $1-\frac{c_{R_{(\mathrm{max},i)}}}{\Lambda_m}$. 

However, for 3D problems, such a local estimate is neither convincing nor practical. A more reliable approach is to minimize the factors in Eq.~(\ref{eq:1dboosttimestep}) \textit{separately}:
\beq\label{eq:boosttimestep}
\Delta t_b = \mathrm{CFL}\min_i\left(\frac{\Delta x_i}{c_{R_{(\mathrm{max},i})}}\right)\min_i\left(1-\frac{c_{R_{(\mathrm{max},i)}}}{\Lambda_m}\right)=\frac{\Delta t}{\Gamma_B},
\eeq
where we used Eq.~(\ref{eq:1Dtimestep}) for the rest frame time step and introduced the \textit{boost factor} (somewhat similar to the Lorentz $\Gamma$-factor):
\beq\label{eq:boostfactor}
\Gamma_B=\frac1{1-\frac{c_{R_{\mathrm{max}}}}{\Lambda_m}}, \qquad c_{R_{\mathrm{max}}}=\max_i\left(c_{R_{(\mathrm{max},i)}}\right).
\eeq
The larger the boost factor (\ie the closer $\Lambda_m$ is to $c_{R_\mathrm{max}}$) is, the longer the maximum allowable forecast time. Eq.~(\ref{eq:boosttimestep}) is convenient to use within any available MHD model/code, since the time step, $\Delta t$, is calculated as usual (as if the simulation is performed in the rest frame), but before using it in the simulation we should divide it by the separately calculated boost factor. 

In a realistic 3D case, the stability of the finite-volume scheme given by Eqs.~(\ref{eq:viatildeU} and \ref{eq:tildeU}) can be shown in the following way.  We assume that the state at the time level, $n+1$, is calculated in two stages. Initially, the approximation, $\bar{U}^{n+1}_i$ for conserved variables at $n+1$ time level is calculated as follows:
\beq
\bar{U}^{n+1}_i =U^n_i-\Gamma_B\frac{\Delta t_b}{\Delta V_i} \sum_j\mathbf{F}_{ij}\cdot\bm{\sigma}_{ij}.
\eeq
This is an explicit scheme that is conditionally stable as long as $\Gamma_B\Delta t_b=\Delta t$ is the time step in the rest frame that satisfies the Courant condition. 
Using the relation, $\tilde{U}=\frac{U}{\Gamma_B}+\frac1{\Lambda_m}\left(c_{R_{\mathrm{max}}}U-\er\cdot\mathbf{F}\right)$, Eqs.~(\ref{eq:viatildeU} and \ref{eq:tildeU}) can be reduced to:
\beq\label{eq:donorcell}
U^{n+1}+
\frac{\Gamma_B}{\Lambda_m}
\left(c_{R_{\mathrm{max}}}U^{n+1}-\er\cdot\mathbf{F}^{n+1}\right)=\bar{U}^{n+1}+\frac{\Gamma_B}{\Lambda_m}
\left(c_{R_{\mathrm{max}}}U^n-\er\cdot\mathbf{F}^n\right).
\eeq
This equation can be interpreted as the implicit donor-cell scheme \cite[see, \eg][]{vanLeer2006}: the fictive cell with the state $\bar{U}^{n+1}$, receives the upwind flux $\frac{\Gamma_B}{\Lambda_m}
\left(c_{R_{\mathrm{max}}}U^n-\er\cdot\mathbf{F}^n\right)$, from the donor cell, in which the state is $U^n$, and donates an implicit flux $\frac{\Gamma_B}{\Lambda_m} \left(c_{R_{\mathrm{max}}}U^{n+1}-\er\cdot\mathbf{F}^{n+1}\right)$ to somewhere. The scheme is stable if $c_{R_{\mathrm{max}}}$ exceeds the maximum perturbation speed in all states, $U^n$, $\bar{U}^{n+1}$, and $U^{n+1}$. The iterative algorithm for solving implicit the Eq.~(\ref{eq:donorcell}) as described in \ref{Subsec:recovery} above survives even if the boost factor, which plays the role of the CFL number for the donor-cell scheme, is as high as $\Gamma_B\sim10-30$. In this case $\frac{c_{R_\mathrm{max}}}{\Lambda_m}\approx0.95$, which allows for a reasonably long forecast time $\Delta t_0\propto\frac1{\Lambda_m}$, even for fast CMEs.

\section{Forecast System Based on the 3D Hydrodynamic Equations}
\label{Sec:hydro}

\subsection{Governing Equations and their Discretization}
\label{Subsec:hydro-disc}
The hydrodynamic motion is governed by the continuity equation, momentum vector conservation law, and energy scalar conservation law, which for a gas with constant $\gamma$ read:
\beq\label{eq:hydro} 
\begin{split}\raisetag{15ex}
&\frac{\partial\rho}{\partial t} = -\divg(\rho\bu), \\
&\frac{\partial(\rho\bu)}{\partial t}= -\divg\left[\rho\bu\bu +p\mathbf{I}\right], \\
&\frac\partial{\partial t}\left(\frac{\rho u^2}2+\frac{p}{\gamma-1}\right)=-\divg\left[\left(\frac{\rho u^2}2+\frac{\gamma p}{\gamma-1}\right)\bu\right].
\end{split} 
\eeq
When applying the finite volume formalism given by Eq.~(\ref{eq:viatildeU}), the numerical fluxes $-\frac{1}{\Delta V_i}  \sum_j\mathbf{F}_{ij}\cdot\bm{\sigma}_{ij}$ on the right-hand side (RHS) of the above equations give us the numerical sources of mass, $S_\rho$, momentum $\mathbf{S}_{\rho\bu}$, and energy $S_{\cal E}$ (herewith, the subscript index, $i$ enumerating the control volumes is omitted). The vectors of velocity and momentum density are conveniently split into radial and ``horizontal'' (perpendicular to the radial direction) components:
\beq\label{eq:Randperp}
u_\sR=\er\cdot\bu,\quad \bu_\perp=\bu-u_\sR\er,\quad S_{\rho u_\sR}=\er\cdot \mathbf{S}_{\rho\bu},\quad \mathbf{S}_{\rho\bu_\perp} =\mathbf{S}_{\rho\bu}-S_{\rho u_\sR}\er.
\eeq

With these definitions Eqs.~(\ref{eq:viatildeU}) for 3D hydrodynamic Eqs.~(\ref{eq:hydro}) give:
\beq\label{eq:discrhydro1}
\begin{split}\raisetag{25ex}
&\tilde{U}_\rho^{n+1} =\Delta t_b\,S_\rho + \tilde{U}^n_\rho, \quad\tilde{U}^n_\rho=\rho^n\left(1-\frac{u^n_\sR}{\Lambda_m}\right)
\\
&\tilde{U}^{n+1}_{\rho u_\sR} =
\Delta t_b S_{\rho u_\sR}
+\tilde{U}^n_{\rho u_\sR},\quad \tilde{U}^n_{\rho u_\sR}=
\rho^n\left(1-\frac{u^n_\sR}{\Lambda_m}\right)u^n_\sR-\frac{p^n}{\Lambda_m},
\\
&\bm{\tilde{U}}^{n+1}_{\rho\bu_\perp} =\Delta t_b\mathbf{S}_{\rho\bu_\perp} +\bm{\tilde{U}}^n_{\rho\bu_\perp},\quad\bm{\tilde{U}}^n_{\rho\bu_\perp} = \rho^n\left(1-\frac{u^n_\sR}{\Lambda_m}\right)\bu^n_\perp,
\\
&\tilde{U}_{\cal E}^{n+1} =\Delta t_b S_{\cal E} + \tilde{U}^n_{\cal E},\quad \tilde{U}^n_{\cal E} =
\frac{\rho^n (\bu^n)^2}2 \left(1-\frac{u^n_\sR}{\Lambda_m}\right) +\frac{p^n}{\gamma-1}\left(1-\frac{\gamma u^n_\sR}{\Lambda_m}\right).
\end{split}
\eeq

\subsection{Solving for Primitive Variables}
\label{Subsec:hydro-primitive}
Let us denote the LHS of the first of Eqs.~(\ref{eq:speeding}) as
\beq\label{eq:densvar}
\bar{\rho}=\rho^{n+1}\left(1-\frac{u^{n+1}_\sR}{\Lambda_m}\right)=\Delta t_b S_\rho+\rho^n\left(1-\frac{u^n_\sR}{\Lambda_m}\right)
\eeq
and use it not only to eliminate the factor $\rho^{n+1}\left(1-u^{n+1}_\sR/\Lambda_m\right) =\bar{\rho}$ from $\tilde{U}^{n+1}_{\rho u_\sR}$, $\bm{\tilde{U}}^{n+1}_{\rho\bu_\perp}$ and $\tilde{U}^{n+1}_{\cal E}$, but also to eliminate the factor, $\rho^n\left(1-u^{n}_\sR/\Lambda_m\right) =\bar{\rho}-\Delta t_b S_\rho$, from $\tilde{U}^n_{\rho u_\sR}$, $\bm{\tilde{U}}^n_{\rho\bu_\perp}$ and $\tilde{U}^n_{\cal E}$:
\beq\label{eq:hydrodiscr2}
\begin{split}
& \tilde{U}^{n+1}_{\rho u_\sR}= \Delta t_b\left(S_{\rho u_\sR}-u_\sR^nS_\rho\right) +\tilde{U}^n_{\rho u_\sR},\quad \tilde{U}^n_{\rho u_\sR}= \bar{\rho} u^n_\sR-\frac{p^n}{\Lambda_m},
\\
&\bm{\tilde{U}}^{n+1}_{\rho\bu_\perp}=\Delta t_b\left(\mathbf{S}_{\rho\bu_\perp}- \bu^n_\perp S_\rho\right)+\bm{\tilde{U}}^n_{\rho\bu_\perp},\quad\bm{\tilde{U}}^n_{\rho\bu_\perp}= \bar{\rho}\bu^n_\perp,
\\
&\tilde{U}_{\cal E}^{n+1}=\Delta t_b\left(S_{\cal E}-\frac{\bu^2}{2}S_{\rho}\right) + \tilde{U}^n_{\cal E},\quad\tilde{U}^n_{\cal E}=
\frac{\bar{\rho} (\bu^n)^2}2+\frac{P^n}{\gamma-1}\left(1-\frac{\gamma u^n_\sR}{\Lambda_m}\right),
\end{split}
\eeq
where we keep notation, $\tilde{U}$, for the \textit{reduced} conserved variables, although the expressions for them are simplified. In the next step, we want to find the primitive variables from Eqs.~(\ref{eq:hydrodiscr2}), \ie the components of the velocity vector and pressure (the density is obtained below from Eq.~(\ref{eq:densvar}), once $u^{n+1}_\sR$ is known). The equation for the perpendicular velocity (the second equation in Eqs.~\ref{eq:hydrodiscr2}) can be solved explicitly: 
\beq\label{eq:uperp}
\bu^{n+1}_\perp=
\frac{\Delta t_b}{\bar{\rho}}
\left(\mathbf{S}_{\rho\bu_\perp}- \bu^n_\perp S_\rho\right)+\bu^n_\perp.
\eeq
Eq.~(\ref{eq:uperp}) can be used to eliminate $\bu^2_\perp$ from the energy equation (the third line in of Eqs.~\ref{eq:hydrodiscr2}), which now becomes:
\begin{eqnarray}
\tilde{U}_{\cal E}^{n+1}&=&\Delta t_b\left[S_{\cal E}-\frac{\bu^2}{2}S_{\rho}-\bu^n_\perp\cdot\left(\mathbf{S}_{\rho\bu_\perp}- \bu^n_\perp S_\rho\right)\right] -\frac{\left(\Delta t_b\right)^2}{2\bar{\rho}}\left(\mathbf{S}_{\rho\bu_\perp}- \bu^n_\perp S_\rho\right)^2+ \tilde{U}^n_{\cal E},\nonumber\\
\tilde{U}^n_{\cal E}&=&
\frac{\bar{\rho} \left(u^n_\sR\right)^2}2+\frac{p^n}{\gamma-1}\left(1-\frac{\gamma u^n_\sR}{\Lambda_m}\right).\label{eq:hydroenergy}
\end{eqnarray}
Finally, let us return to the radial component of the velocity, described by the first line in Eqs.~(\ref{eq:hydrodiscr2}). This equation can be considered as a linear relation between $\delta u_\sR=u^{n+1}-u^n_\sR$ and $\delta p = p^{n+1}-p^n$:
where:
\beq\label{eq:hydrolin}
\delta u_\sR = \frac{\Delta t_b}{\bar{\rho}}\left(S_{\rho u_\sR}-u_\sR^nS_\rho\right)+\frac{\delta p}{\bar{\rho}\Lambda_m}.
\eeq
Eliminating $\delta u_\sR$ from Eq.~Eq.~(\ref{eq:hydrodiscr2}) with the use of Eq.~(\ref{eq:hydrolin}) results in a quadratic equation, 
\beq\label{eq:deltap}
\alpha\left(\delta p\right)^2-\beta\delta p+\varepsilon=0
\eeq
that can be solved
\beq\label{eq:delta-p}
\delta p= \frac{2\varepsilon}{\beta+\sqrt{\beta^2-4\alpha\varepsilon}},
\eeq
where
\begin{eqnarray}
&&\alpha=\frac{\left(\gamma+1\right)}{2\left(\gamma-1\right)\bar{\rho}\Lambda^2_m},\quad \beta=\frac{1}{\gamma-1}\left(1-\frac{u^n_\sR}{\Lambda_m}-\frac{\Delta t_b\left(S_{\rho u_\sR}-u_\sR^nS_\rho\right)}{\bar{\rho}\Lambda_m}-\frac{\gamma p^n}{\bar{\rho}\Lambda^2_m}\right),\\
&&\varepsilon=\Delta t_b\left[S_{\cal E}+\frac{\bu^2}{2}S_{\rho}-\bu^n\cdot\mathbf{S}_{\rho\bu}+\frac{\gamma p^n\left(S_{\rho u_\sR}-u_\sR^nS_\rho\right)}{\left(\gamma-1\right)\bar{\rho}\Lambda_m} \right] -\frac{\left(\Delta t_b\right)^2}{2\bar{\rho}}\left(\mathbf{S}_{\rho\bu}- \bu^n S_\rho\right)^2.\nonumber
\end{eqnarray}
We note that formally $\delta p=2\varepsilon/(\beta-\sqrt{\beta^2-4\alpha\varepsilon})$ is also a solution of Eq.~(\ref{eq:deltap}). However, for small $\varepsilon\propto\Delta t_b$ it is not a physically meaningful solution as it does not tend to zero as $\Delta t_b\rightarrow 0$, while Eq.~(\ref{eq:delta-p}) automatically satisfies the correct asymptotic expansion, $\delta p = \varepsilon/\beta+O[\varepsilon^2]$.

With $\delta p$ obtained from Eq.~(\ref{eq:delta-p}) one can find $p^{n+1}=p^n+\delta p$ and solve for $\delta u_\sR$ from Eq.~(\ref{eq:hydrolin}), then solve $u^{n+1}_\sR=u^n_\sR+\delta u_\sR$ and derive the density from Eq.~(\ref{eq:densvar}): $\rho^{n+1}=\bar{\rho}/\left(1-u^{n+1}_\sR/\Lambda_m\right)$. In this way, all primitive variables are recovered.

\subsection{Numerical Hydrodynamics Test}
\label{Subsec:numerical1}

\begin{figure}[ht!]
\centering
\includegraphics[width=0.495\linewidth]{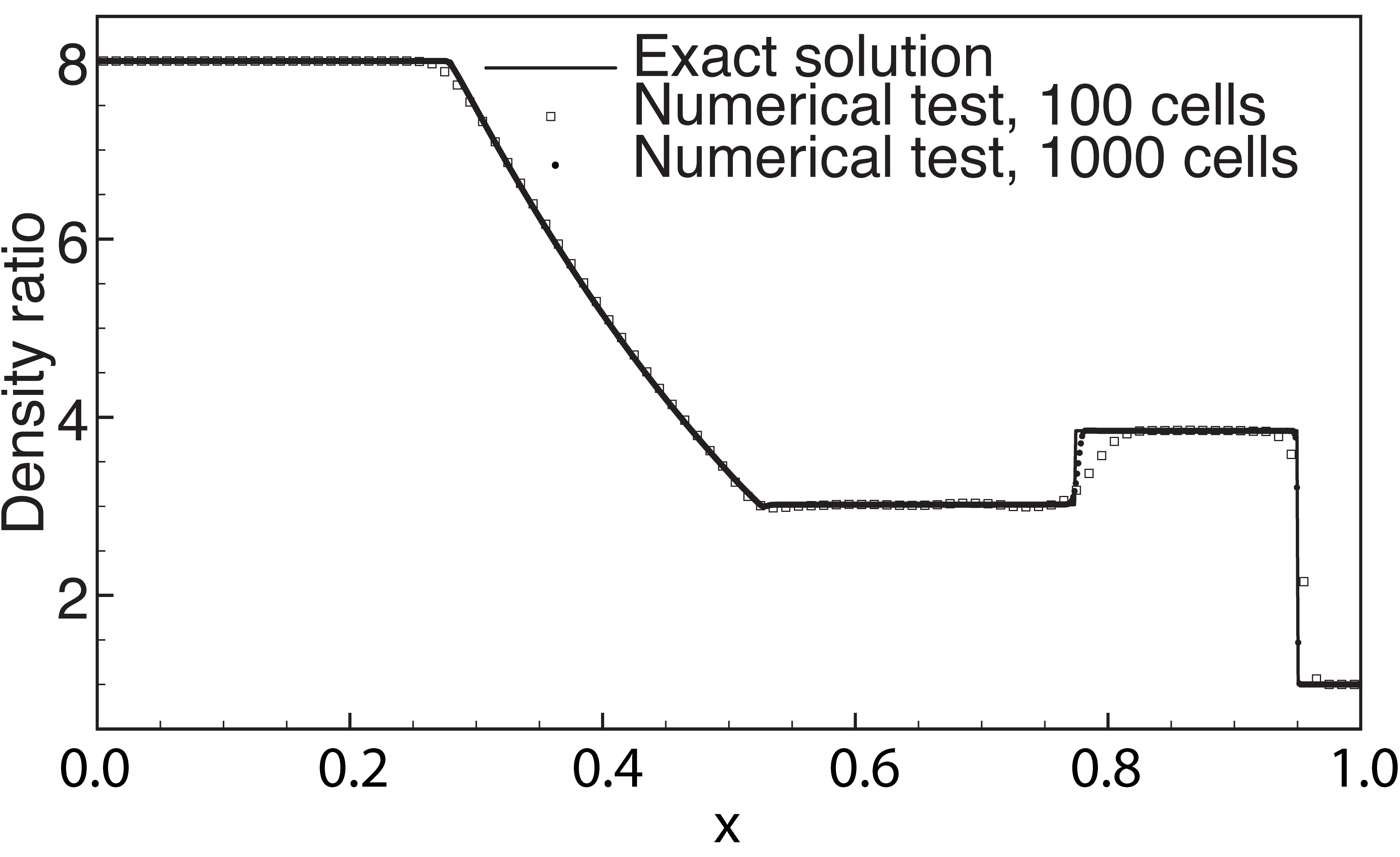}
\includegraphics[width=0.495\linewidth]{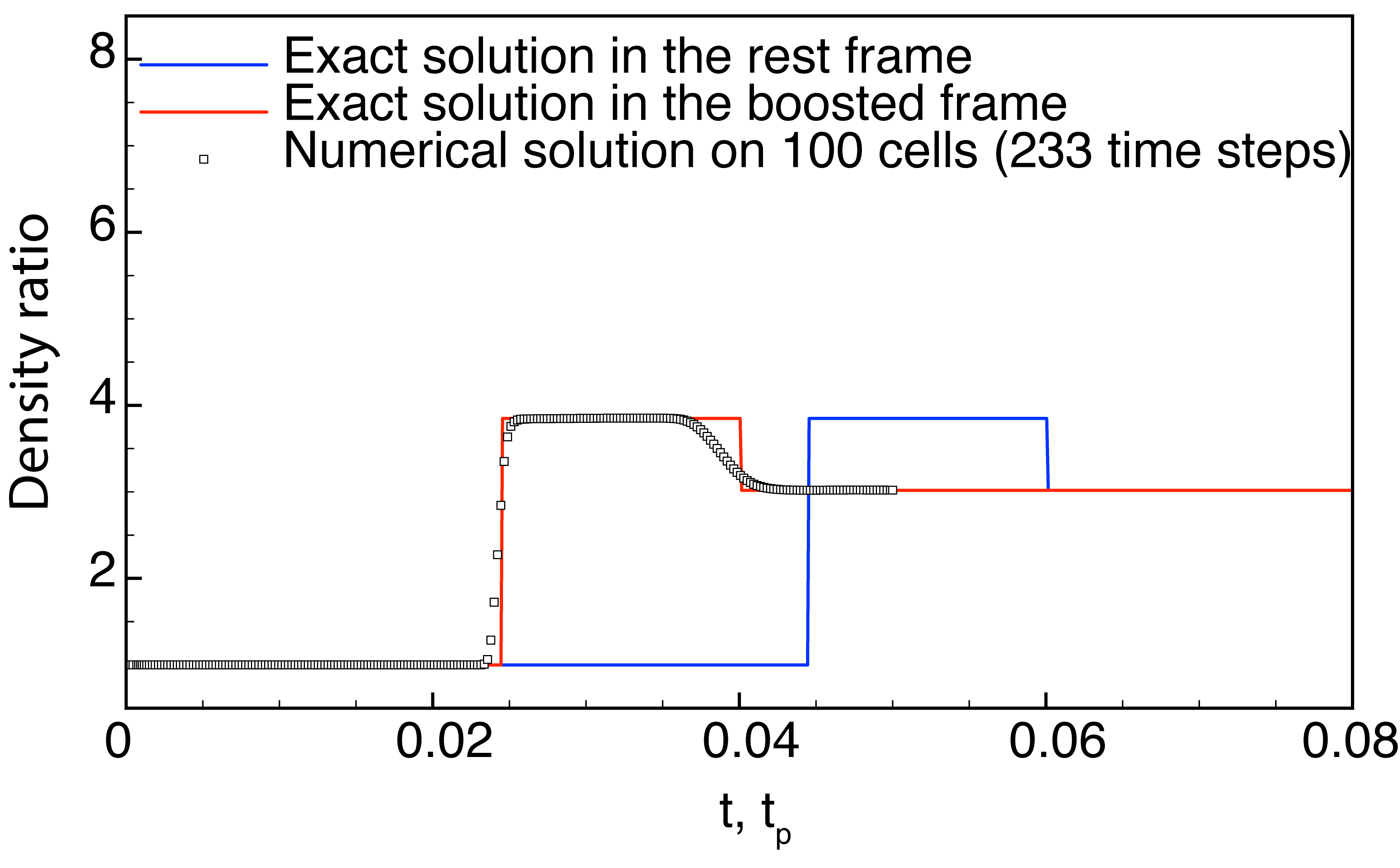}
\caption{Solution of the Riemann problem for a gas with $\gamma=5/3$ with the initial condition: $\rho=8$, $u=0$ and $p=480$ for $x<0.5$; $\rho=1$, $u=0$ and $p=1$ for $x>0.5$. {\bf Left panel:} Density ratio as a function of $x$ in the boosted frame at time $t_b=0.022$: the solid line shows the exact solution (also shown with the red line in Fig.~\ref{fig:scheme}); with symbols the numerical test solution is presented on 100 and 1000 cells.  {\bf Right panel:} Time profile of density at $x=1$. Red line is the solution in the boosted frame, blue line is the solution in the rest frame. $\rho(t_b)$ predicts the solution in the rest frame, $\rho(t)$ with the time offset of $0.02$ (both lines are the same as those in Fig.~(\ref{fig:scheme}); symbols show the test simulation results at 233 time steps, with 100 cells per an interval, $(0,1)$, of x.}
\label{fig:testhydro}
\end{figure}
To test the proposed scheme, we solved numerically the Riemann problem with discontinuous initial condition as described in Section~\ref{Sec:characteristic} and illustrated in Fig.~\ref{fig:scheme} and compare the obtained numerical results with the exact solutions provided there. The numerical fluxes are calculated on the basis of the exact Riemann solver (the \textit{Godunov scheme}) as described and coded by \cite{Toro2009}. To achieve the second order of accuracy, a \textit{limited interpolation procedure} is applied to obtain left and right interpolated states of primitive variables on each face of the control volumes to be used as inputs for the numerical flux. The two-stage Runge-Kutta scheme is used for time integration.

In our numerical example, the interval $(0,1)$ of the 1D coordinate, $x$, is divided into 100 equally spaced cells with a mesh size of 0.01 and an initial discontinuity is located at x=0.5. To the right from this location, the boosted frame is used in a way that the time offset at the right boundary reaches $0.02$, corresponding to a limiting speed of $\Lambda_m =0.5/0.02=25$. In Fig.~\ref{fig:testhydro} the top panel shows the density distribution over $x$ at time $0.022$.  The simulation took 105 time steps with a Courant-Friedrichs-Levi (CFL) number of 0.9. The agreement with the exact solution is reasonable with 100 cells and becomes nearly perfect with 1000 cells (1015 time steps). In the bottom panel, the time evolution of the density at $x=1$ is presented. The blue line (rest frame) and the red line (boosted frame) are identical to those provided in the right panel of Fig.~\ref{fig:scheme}) and show the exact solution. By symbols we present the numerical test result. Keeping the same spatial resolution of 0.01, \ie using 100 cells over the interval of $(0,1)$, we extend the computational domain to $(-0.5,1.5)$, to reduce the effect from the boundaries of the domain. For the simulation time interval, $(0,0.05)$, we show results after 233 time steps obtained in the boosted frame. The test demonstrates a reasonable quality of computational results and can serve as a proof of principle of the proposed forecast system: a numerical solution with time offset is possible.

\section{Numerical Algorithm for Solving the Boosted MHD Equations}
\label{Sec:numericalmhd}
%
To solve Eq.~(\ref{eq:mhd1}) numerically, the vectors of magnetic field, velocity and their fluxes can be conveniently split into the radial velocity/field and the ``horizontal'' components, perpendicular to the radial direction:
\beq
B_\sR=\er\cdot\bB,\quad \bB_\perp=\bB-B_\sR\er,\quad u_\sR=\er\cdot\bu,\quad \bu_\perp=\bu-u_\sR\er,\quad etc.
\eeq
With these transformations, the governing equations become:
\beq\label{eq:mhd2} 
\begin{split}\raisetag{25ex}
&\frac{\partial}{\partial t_b}\left[\rho\left(1-\frac{u_\sR}{\Lambda_m}\right)\right]=S_\rho\\
&\frac{\partial}{\partial t_b}\left[\rho\left(1-\frac{u_\sR}{\Lambda_m}\right)u_\sR-\frac1{\Lambda_m}\left(P+\frac{B_\perp^2}{2\mu_0}\right)\right]=S_{\rho u_\sR}\\
&\frac{\partial}{\partial t_b}\left[\rho\left(1-\frac{u_\sR}{\Lambda_m}\right)\bu_\perp\right]+\frac{B_\sR}{\Lambda_m\mu_0}\frac{\partial\bB_\perp}{\partial t_b}=\mathbf{S}_{\rho\bu_\perp}\\
&\left(1-\frac{u_\sR}{\Lambda_m}\right)\frac{\partial B_\sR}{\partial t_b}=S_{B_\sR}\\
&\frac{\partial}{\partial t_b}\left[\left(1-\frac{u_\sR}{\Lambda_m}\right)\bB_\perp\right]+\frac{B_\sR}{\Lambda_m}\frac{\partial\bu_\perp}{\partial t_b}=\mathbf{S}_{\bB_\perp}
\end{split} 
\eeq
The energy equation,
\begin{eqnarray}
&&\frac{\partial}{\partial t_b} \left[\frac{\rho u^2\left(1-\frac{u_\sR}{\Lambda_m}\right)}2+\frac{P\left(1-\frac{\gamma u_\sR}{\Lambda_m}\right)}{\gamma-1}+\frac{B^2_\perp\left(1-\frac{2u_\sR}{\Lambda_m}\right)+B_\sR^2}{2\mu_0}\right]+\nonumber\\
&&\quad+\frac{B_\sR}{\Lambda_m\mu_0}\left[\frac{\partial\left(\bu_\perp\cdot\bB_\perp\right)}{\partial t_b}-u_\sR\frac{\partial B_\sR}{\partial t_b}\right]=S_{\cal E},\nonumber
\end{eqnarray}
can be obtained in a more convenient form  by multiplying the radial field equation by $B_\sR/\mu_0$ and subtracting the result from the energy equation:
\beq\label{eq:energy-a}
\frac\partial{\partial t_b}\left[\frac{\rho u^2\left(1-\frac{u_\sR}{\Lambda_m}\right)}2+\frac{P\left(1-\frac{\gamma u_\sR}{\Lambda_m}\right)}{\gamma-1}+\frac{B^2_\perp\left(1-\frac{2u_\sR}{\Lambda_m}\right)}{2\mu_0}\right]+\frac{B_\sR}{\Lambda_m\mu_0}\frac{\partial\left(\bu_\perp\cdot\bB_\perp\right)}{\partial t_b}=S_{\cal E}-\frac{B_\sR}{\mu_0}S_{B_\sR}.
\eeq

\subsection{Algorithm}
\label{Subsec:algorithm}
Within the framework of finite volume numerical methods, the governing equations (Eqs.~\ref{eq:mhd2}) can be integrated over control volume using Gauss' theorem, then the RHS of the equations are calculated as the sum of \textit{numerical fluxes} through the faces of the \textit{control volume}, divided by the magnitude of control volume. Once the numerical solution is known at the instant of time $t_b=t^n$, and the numerical fluxes and sources in the RHS of Eqs.~(\ref{eq:mhd2}) are all calculated, the solution at the next time level, $t_b=t^{n+1}=t^n+dt$ can be found by integrating the equations over time, giving:
\beq\label{eq:discr1}
\begin{split}\raisetag{25ex}
& \tilde{U}_\rho^{n+1} =\Delta t_b S_\rho + \tilde{U}^n_\rho, \quad\tilde{U}^n_\rho=\rho^n\left(1-\frac{u^n_\sR}{\Lambda_m}\right)
\\
& \tilde{U}^{n+1}_{\rho u_\sR} =
\Delta t_b S_{\rho u_\sR}
+\tilde{U}^n_{\rho u_\sR},\quad \tilde{U}^n_{\rho u_\sR}=
\rho^n\left(1-\frac{u^n_\sR}{\Lambda_m}\right)u^n_\sR-\frac1{\Lambda_m}\left(p^n+\frac{(\bB^n_\perp)^2}{2\mu_0}\right),
\\
& \bm{\tilde{U}}^{n+1}_{\rho\bu_\perp} =\Delta t_b\mathbf{S}_{\rho\bu_\perp} +\bm{\tilde{U}}^n_{\rho\bu_\perp},\quad\bm{\tilde{U}}^n_{\rho\bu_\perp} = \rho^n\left(1-\frac{u^n_\sR}{\Lambda_m}\right)\bu^n_\perp+\frac{\bar{B}_\sR}{\Lambda_m\mu_0}\bB^n_\perp,
\\
& \left(1-\frac{\bar{u}_\sR}{\Lambda_m}\right)B^{n+1}_\sR =\Delta t_b S_{B_\sR}+\left(1-\frac{\bar{u}_\sR}{\Lambda_m}\right)B^n_\sR
\\
& \bm{\tilde{U}}^{n+1}_{\bB_\perp} =
\Delta t_b\mathbf{S}_{\bB_\perp}+
\bm{\tilde{U}}^n_{\bB_\perp},\quad 
\bm{\tilde{U}}^n_{\bB_\perp} =
\left(1-\frac{u^n_\sR}{\Lambda_m}\right)\bB^n_\perp+\frac{\bar{B}_\sR}{\Lambda_m}\bu^n_\perp
\\
& \tilde{U}_{\cal E}^{n+1} =\Delta t_b\left(S_{\cal E} -\frac{\bar{B}_\sR}{\mu_0}S_{B_\sR}\right) + \tilde{U}^n_{\cal E},\\
& \tilde{U}^n_{\cal E} =
\frac{\rho^n \left(\bu^n\right)^2}2 \left(1-\frac{u^n_\sR}{\Lambda_m}\right) +\frac{p^n\left(1-\frac{\gamma u^n_\sR}{\Lambda_m}\right)}{\gamma-1} +\frac{\left(\bB^n_\perp\right)^2 \left(1-\frac{2u^n_\sR}{\Lambda_m}\right)}{2\mu_0} +\frac{\bar{B}_\sR}{\Lambda_m\mu_0}\left(\bu^n_\perp\cdot\bB^n_\perp\right).
\end{split}
\eeq
The coefficient, $\bar{u}_\sR$, multiplied by the time derivative, $\partial B_\sR/\partial t_b$ in the field equations in Eqs.~(\ref{eq:mhd2}) is present in both the $B^n_\sR$ and $B^{n+1}_\sR$ states in Eqs.~(\ref{eq:discr1}). It originates from the explicit source term, $-u_\sR\divg\bB$, contributing to $S_{B_\sR}$, therefore, it is known when Eqs.~(\ref{eq:discr1}) needs to be solved. Therefore, $B^{n+1}_\sR$, can be solved explicitly. For another coefficient, $\bar{B}_\sR$, which also stands out of time derivatives and is present in both $n$th and $(n+1)$th states, may be also solved explicitly as $\bar{B}_\sR=\frac12\left(B^{n+1}_\sR+B^n_\sR\right)$:
\beq\label{eq:BrNewandBar}
B^{n+1}_\sR=B^n_\sR+\frac{\Delta t_b S_{B_\sR}}{1-\frac{\bar{u}_\sR}{\Lambda_m}},\qquad \bar{B}_\sR=B^n_\sR+\frac{\Delta t_b S_{B_\sR}}{2\left(1-\frac{\bar{u}_\sR}{\Lambda_m}\right)}.
\eeq
To simplify the algorithm, we follow the steps described in Appendix \ref{Subsec:hydro-primitive}. First, we use Eq.~(\ref{eq:densvar}) to eliminate the densities, $\rho^n$, $\rho^{n+1}$, from the expressions for $\tilde{U}_{\rho u_\sR}$, $\bm{\tilde{U}}_{\rho\bu_\perp}$ \text{and} $\tilde{U}_{\cal E}$.
Now, the equations for these conserved variables are:
\beq\label{eq:momentumhdiscr2}
\begin{split}
& \tilde{U}^{n+1}_{\rho u_\sR}= \Delta t_b\left(S_{\rho u_\sR}-u_\sR^nS_\rho\right) +\tilde{U}^n_{\rho u_\sR},\quad \tilde{U}^n_{\rho u_\sR}= \bar{\rho} u^n_\sR-\frac1{\Lambda_m}\left(p^n+\frac{(\bB^n_\perp)^2}{2\mu_0}\right),
\\
&\bm{\tilde{U}}^{n+1}_{\rho\bu_\perp}=\Delta t_b\left(\mathbf{S}_{\rho\bu_\perp}- \bu^n_\perp S_\rho\right)+\bm{\tilde{U}}^n_{\rho\bu_\perp},\quad\bm{\tilde{U}}^n_{\rho\bu_\perp}= \bar{\rho}\bu^n_\perp+\frac{\bar{B}_\sR}{\Lambda_m\mu_0}\bB^n_\perp,
\\
&\tilde{U}_{\cal E}^{n+1}=\Delta t_b\left(S_{\cal E}-\frac{\left(\bu^n\right)^2}2S_\rho-\frac{\bar{B}_\sR}{\mu_0}S_{B_\sR}\right) + \tilde{U}^n_{\cal E},
\\
&\tilde{U}^n_{\cal E}=
\frac{\bar{\rho} \left(\bu^n\right)^2}2+\frac{p^n\left(1-\frac{\gamma u^n_\sR}{\Lambda_m}\right)}{\gamma-1}+\frac{\left(\bB^n_\perp\right)^2\left(1-\frac{2u^n_\sR}{\Lambda_m}\right)}{2\mu_0}+\frac{\bar{B}_\sR}{\Lambda_m\mu_0}\left(\bu^n_\perp\cdot\bB^n_\perp\right).
\end{split}
\eeq
Next, we express $\bu^{n+1}_\perp$ in terms of the yet unknown $\bB^{n+1}_\perp$:
\beq\label{eq:solveuperp}
\bu^{n+1}_\perp=
\frac{\Delta t_b}{\bar{\rho}}
\left(\mathbf{S}_{\rho\bu_\perp}- \bu^n_\perp S_\rho\right)
+\bu^n_\perp-
\frac{\bar{B}_\sR}{\Lambda_m\mu_0\bar{\rho}}\left(\bB^{n+1}_\perp-\bB^{n}_\perp\right)
\eeq
and then use Eq.~(\ref{eq:solveuperp}) to eliminate $\bu^{n+1}_\perp$ from the equations for the energy and the magnetic field. 

Finally, we obtain the finite-volume scheme for the reduced conserved variables:
\beq\label{eq:mhddiscr2}
\begin{split}
& \tilde{U}^{n+1}_{\rho u_\sR}= 
\Delta t_b\left(S_{\rho u_\sR}-u_\sR^nS_\rho\right) 
+\tilde{U}^n_{\rho u_\sR},\quad 
\tilde{U}^n_{\rho u_\sR}= \bar{\rho} u^n_\sR-\frac1{\Lambda_m}\left(p^n+\frac{(\bB^n_\perp)^2}{2\mu_0}\right),
\\
& \bm{\tilde{U}}^{n+1}_{\bB_\perp} =\Delta t_b\left[\mathbf{S}_{\bB_\perp}
-\frac{\bar{B}_\sR}{\bar{\rho}\Lambda_m}
\left(\mathbf{S}_{\rho\bu_\perp}- \bu^n_\perp S_\rho\right)\right]+\bm{\tilde{U}}^n_{\bB_\perp}
,\quad 
\bm{\tilde{U}}^n_{\bB_\perp} =
\left(1-\frac{u^n_\sR}{\Lambda_m}-\frac{\bar{V}^2_{\sA_\sR}}{\Lambda_m^2}\right)\bB^n_\perp,
\\
&\tilde{U}_{\cal E}^{n+1}=\Delta t_b\left[S_{\cal E}-\frac{\left(\bu^n\right)^2}2S_\rho-\left(\bu^n_\perp+\frac{\bar{B}_\sR\bB^n_\perp}{\mu_0\Lambda_m\bar{\rho}}\right)\cdot\left(\mathbf{S}_{\rho\bu_\perp}- \bu^n_\perp S_\rho\right)-\frac{\bar{B}_\sR}{\mu_0}S_{B_\sR}\right]-\\
&\quad-\frac{\left(\Delta t_b\right)^2\left(\mathbf{S}_{\rho\bu_\perp}- \bu^n_\perp S_\rho\right)^2}{2\bar{\rho}} + \tilde{U}^n_{\cal E},\\
&\tilde{U}^n_{\cal E}=
\frac{\bar{\rho} (u_\sR^n)^2}2+\left(1-\frac{\gamma u^n_\sR}{\Lambda_m}\right)\frac{p^n}{\gamma-1}+\left(1-\frac{2u^n_\sR}{\Lambda_m}-\frac{\bar{V}^2_{\sA_\sR}}{\Lambda_m^2}\right)\frac{\left(\bB^n_\perp\right)^2}{2\mu_0},
\end{split}
\eeq
where $\bar{V}_{\sA_\sR}^2 = \bar{B}_\sR^2/(\mu_0\bar{\rho})$ is the square of the \alf~wave speed in the radial direction.

\subsection{Solving for Primitive Variables}
\label{Subsec:primitive}
Eqs.~(\ref{eq:mhddiscr2}) can be solved and the unknown $\bm{{\cal P}} =\left(u^{n+1}_\sR, \bB^{n+1}_\perp,p^{n+1}\right)$ can be found using the method described in \ref{Subsec:recovery}. The conserved variables, $\tilde{U}^{(n+1,h)}$, in iteration $h$ can be expressed in terms of the primitive variables, $\bm{{\cal P}}^{(n+1,h)} =\left(u^{(n+1,h)}_\sR, \bB^{(n+1,h)}_\perp, p^{(n+1,h)}\right)$:
\beq\label{eq:mhddiscr3}
\begin{split}
& \tilde{U}_{\rho u_\sR}^{(n+1,h)}= \bar{\rho} u^{(n+1,h)}_\sR-\frac1{\Lambda_m}\left[p^{(n+1,h)}+\frac{\left(\bB^{(n+1,h)}_\perp\right)^2}{2\mu_0}\right],
\\
& \bm{\tilde{U}}_{\bB_\perp}^{(n+1,h)} =
\left(1-\frac{u^{(n+1,h)}_\sR}{\Lambda_m}-\frac{\bar{V}^2_{\sA_\sR}}{\Lambda_m^2}\right)\bB^{(n+1,h)}_\perp,
\\
&\tilde{U}_{\cal E}^{(n+1,h)}=
\frac{\bar{\rho} \left(u_\sR^{(n+1,h)}\right)^2}2+
\left(1-\frac{\gamma u^{(n+1,h)}_\sR}{\Lambda_m}\right)
\frac{p^{(n+1,h)}}{\gamma-1}+
\left(1-\frac{2u^{(n+1,h)}_\sR}{\Lambda_m}-
\frac{\bar{V}^2_{\sA_\sR}}{\Lambda_m^2}\right)
\frac{\left(\bB^{(n+1,h)}_\perp\right)^2}{2\mu_0}.
\end{split}
\eeq
At time level $n+1$ we start the iteration process ($h=0$) from quantities obtained at the previous time level, $t^n$:  $\bm{{\cal P}}^{(n+1,0)}=\bm{{\cal P}}^{n}$. We use Eqs.~(\ref{eq:recover}) to solve the increments of the primitive variables, $\delta^{h+1}\bm{{\cal P}}=\bm{{\cal P}}^{(n+1,h+1)}-\bm{{\cal P}}^{(n+1,h)}$. We linearize the increments of the conserved variables using Eq.~(\ref{eq:linearized}):
\beq\label{eq:mhddiscr4}
\begin{split}
& \tilde{U}_{\rho u_\sR}^{(n+1,h+1)}-\tilde{U}_{\rho u_\sR}^{(n+1,h)} \approx \bar{\rho}\,\delta^{h+1}u_\sR-\frac{\bB^{(n+1,h)}_\perp}{\mu_0\Lambda_m}\cdot\delta^{h+1}\bB_\perp-\frac1{\Lambda_m}\delta^{h+1}p,
\\
& \bm{\tilde{U}}_{\bB_\perp}^{(n+1,h+1)}-\bm{\tilde{U}}_{\bB_\perp}^{(n+1,h)} \approx
-\frac{\bB^{(n+1,h)}_\perp}{\Lambda_m}\delta^{h+1}u_\sR+
\left(1-\frac{u^{(n+1,h)}_\sR}{\Lambda_m}-\frac{\bar{V}^2_{\sA_\sR}}{\Lambda_m^2}\right)\delta^{h+1}\bB_\perp,
\\
&\tilde{U}_{\cal E}^{(n+1,h+1)}-\tilde{U}_{\cal E}^{(n+1,h)}-u^{(n+1,h)}_\sR\left(\tilde{U}_{\rho u_\sR}^{(n+1,h+1)}-\tilde{U}_{\rho u_\sR}^{(n+1,h)}\right)-\\
&-\frac{\bB^{(n+1,h)}_\perp}{\mu_0}\cdot\left(\bm{\tilde{U}}_{\bB_\perp}^{(n+1,h+1)}-\bm{\tilde{U}}_{\bB_\perp}^{(n+1,h)}\right)\approx-\frac{\gamma p^{(n+1,h)}}{\left({\gamma-1}\right)\Lambda_m}\delta^{h+1}u_\sR +\frac{1-u^{(n+1,h)}_\sR/\Lambda_m}{\gamma-1}
\delta^{h+1}p,
\end{split}
\eeq
Here we introduced the increment of the (non-conserved) \textit{internal energy} in the last equation instead of the increment in the full energy. This helps us to simplify the last (and quite cumbersome) equation in Eqs.~(\ref{eq:recover}) by subtracting a reasonably chosen linear combination of the other equations. The iterated energy defects in the RHS of Eqs.~(\ref{eq:recover}) can be derived from Eqs.~(\ref{eq:endefect} and \ref{eq:mhddiscr3}):
\beq\label{eq:endefect1}
\begin{split}
\Delta^h_{\rho u_\sR} &= \left\{
\renewcommand*{\arraystretch}{1.75}
\begin{array}{ll}
\Delta t_b\left(S_{\rho u_\sR}-u_\sR^nS_\rho\right), & h=0 \\
\Delta^{h-1}_{\rho u_\sR}+\tilde{U}_{\rho u_\sR}^{(n+1,h-1)} -\tilde{U}_{\rho u_\sR}^{(n+1,h)}, & h\ge1 \\
\end{array}
\right.
\\
\bm{\Delta}^h_{\bB_\perp} &=\left\{
\renewcommand*{\arraystretch}{1.75}
\begin{array}{ll}
\Delta t_b\left[\mathbf{S}_{\bB_\perp}-\frac{\bar{B}_\sR}{\bar{\rho}\Lambda_m}\left(\mathbf{S}_{\rho\bu_\perp}- \bu^n_\perp S_\rho\right)\right] , &h=0\\
\bm{\Delta}^{h-1}_{\bB_\perp}+\bm{\tilde{U}}_{\bB_\perp}^{(n+1,h-1)} -\bm{\tilde{U}}_{\bB_\perp}^{(n+1,h)}, &h\ge1
\end{array}
\right.
\\
\Delta^h_{\cal E} &= \left\{
\renewcommand*{\arraystretch}{1.75}
\begin{array}{ll}
\Delta t_b\left[S_{\cal E}+\frac{\left(\bu^n\right)^2}2S_\rho-\bu^n\cdot\mathbf{S}_{\rho\bu}-\frac{\bar{B}_\sR}{\mu_0}S_{B_\sR}-\frac{\bB^n_\perp}{\mu_0}\cdot\mathbf{S}_{\bB_\perp}\right] &\\
\hspace{2.5em} -\frac{\left(\Delta t_b\right)^2\left(\mathbf{S}_{\rho\bu_\perp}- \bu^n_\perp S_\rho\right)^2}{2\bar{\rho}}
, &h=0\\
\bm{\Delta}^{h-1}_{\cal E}+u^{(n+1,h-1)}_\sR\Delta^{h-1}_{\rho u_\sR} +\frac{\bB^{(n+1,h-1)}_\perp}{\mu_0}\cdot\bm{\Delta}^{h-1}_{\bB_\perp} \\
\hspace{2.5em} +\left[\tilde{U}_{\cal E}^{(n+1,h-1)}-\tilde{U}_{\cal E}^{(n+1,h)}\right] -u^{(n+1,h)}_\sR\Delta^{h}_{\rho u_\sR} -\frac{\bB^{(n+1,h)}_\perp}{\mu_0}\cdot\bm{\Delta}^h_{\bB_\perp}. &h\ge1
\end{array}
\right.
\renewcommand*{\arraystretch}{1.}
\end{split}
\eeq 
Note that in the energy equation we moved into the second line the higher order terms. Solving the system of three linear equations, Eq.~(\ref{eq:recover}) with the LHS derived from Eqs.~(\ref{eq:linearized} and \ref{eq:mhddiscr4})
\beq\label{eq:deltah+1}
\left(\begin{array}{ccc}
\bar{\rho}&-\frac{\bB^{(n+1,h)}_\perp}{\mu_0\Lambda_m}&-\frac1{\Lambda_m}\\
-\frac{\bB^{(n+1,h)}_\perp}{\Lambda_m}&
\left(1-\frac{u^{(n+1,h)}_\sR}{\Lambda_m}-\frac{\bar{V}^2_{\sA_\sR}}{\Lambda_m^2}\right)\mathbf{I}&0\\
-\frac{\gamma p^{(n+1,h)}}{\left({\gamma-1}\right)\Lambda_m}&0&\frac{1-u^{(n+1,h)}_\sR/\Lambda_m}{\gamma-1}
\end{array}\right)\cdot
\left(\begin{array}{c}
\delta^{h+1}u_\sR\\
\delta^{h+1}\bB_\perp\\
\delta^{h+1}p
\end{array}\right)=\left(\begin{array}{c}
\Delta^h_{\rho u_\sR}\\
\bm{\Delta}^h_{\bB_\perp}\\
\Delta^h_{\cal E}
\end{array}\right)
\eeq
is straightforward and can be generalized for many extensions of the MHD model. All increments in the primitive variables are expressed in terms of $\delta^{h+1}u_R$:
\beq\label{eq:alldelta}
\left(\begin{array}{c}
\delta^{h+1}\bB_\perp\\
\delta^{h+1}p
\end{array}\right)=\mathbf{D}^{-1}\cdot\left[\left(\begin{array}{c}
\bm{\Delta}^h_{\bB_\perp}\\
\Delta^h_{\cal E}
\end{array}\right)-\mathbf{A}_{\bm{{\cal P}},u_\sR}\delta^{h+1}u_\sR\right]
\eeq
where 
\beq\label{eq:Dmatrix}
\mathbf{D }=\mathrm{diag}\left[
\left(
1-\frac{u^{(n+1,h)}_\sR}{\Lambda_m}-\frac{\bar{V}^2_{\sA_\sR}}{\Lambda_m^2}\right)\mathbf{I},\frac{1-u^{(n+1,h)}_\sR/\Lambda_m}{\gamma-1}\right]
\eeq 
is an easy-to-invert 4*4 diagonal matrix, $\mathbf{D}^{-1}$ is its inverse, and 
\beq\label{eq:Apu}
\mathbf{A}_{\bm{{\cal P}},u_\sR}=\left(
-\frac{\bB^{(n+1,h)}_\perp}{\Lambda_m},
-\frac{\gamma p^{(n+1,h)}}{\left({\gamma-1}\right)\Lambda_m}\right)^T
\eeq
is a vector of the derivatives of the conserved variables with respect of $u_\sR$. The solution for $\delta^{h+1}u_\sR$ is then given by the first of Eqs.~(\ref{eq:deltah+1}):
\beq\label{eq:deltau}
\delta^{h+1}u_\sR=\frac{\Delta^h_{\rho u_\sR} -\mathbf{A}_{u_R,\bm{{\cal P}}}\cdot\mathbf{D}^{-1}\cdot\left(\begin{array}{c}
\bm{\Delta}^h_{\bB_\perp}\\
\Delta^h_{\cal E}
\end{array}\right) }{\bar{\rho}-\mathbf{A}_{u_R,\bm{{\cal P}}}\cdot\mathbf{D}^{-1}\cdot \mathbf{A}_{\bm{{\cal P}},u_R}}.
\eeq
Here 
\beq\label{eq:Aup}
\mathbf{A}_{u_R,\bm{{\cal P}}}=\left(-\frac{\bB^{(n+1,h)}_\perp}{\mu_0\Lambda_m},-\frac1{\Lambda_m}\right)^T
\eeq
is a vector of the derivatives of the radial stress over the primitive variables multiplied by a factor of $-1/\Lambda_m$. After some algebra one can prove that if in all iterated states, $\bm{{\cal P}}^{(n+1,h)}$, the fast magnetosonic speed in the radial direction does not exceed $\Lambda_m$, then both the matrix $\mathbf{D}$ is positive definite (hence, non-singular) and the denominator in Eq.~(\ref{eq:deltau}) is positive, thus ensuring the convergence of the iteration procedure.

For systems that are more complex than MHD with a single scalar pressure (such more complex systems can also include electron pressure, \alf~wave turbulence, etc), the basic structure of Eq.~(\ref{eq:deltau}) that gives the solution for $\delta^{h+1}u_\sR$ remains the same and only the length of vectors in Eqs.~(\ref{eq:Apu} and \ref{eq:Aup}) increases. 

With known $\delta^{h+1}u_\sR$, the other increments of primitive variables can be solved from Eqs.~(\ref{eq:alldelta}) and then a new iteration for primitive variables can be found, $\bm{{\cal P}}^{(n+1,h+1)}= \bm{{\cal P}}^{(n+1,h)}+\delta^{h+1}\bm{{\cal P}}$. 
If the desired accuracy has not been achieved, the RHS in Eqs.~(\ref{eq:deltah+1}) can be iterated using Eqs.~(\ref{eq:endefect1}); otherwise, the (h+1)st iteration is assigned to the time level, $t^{n+1}$: $\bm{{\cal P}}^{(n+1,h+1)}\rightarrow  \bm{{\cal P}}^{n+1}$ for radial velocity, perpendicular magnetic field and pressure.
The radial magnetic field should be obtained from Eq.~(\ref{eq:BrNewandBar}). 
Once $\bB_\perp^{n+1}$ is known, the perpendicular velocity can be found from Eq.~(\ref{eq:solveuperp}). 
Finally, the density is solved from Eq.~(\ref{eq:densvar}): $\rho^{n+1} =\bar{\rho}/\left(1-u^{n+1}_\sR/\Lambda_m\right)$. In this way, all primitive variables are recovered.

\subsection{Numerical Example}
\label{Subsec:brio-wu}

\begin{figure}[ht!]
\centering
\includegraphics[width=0.495\linewidth]{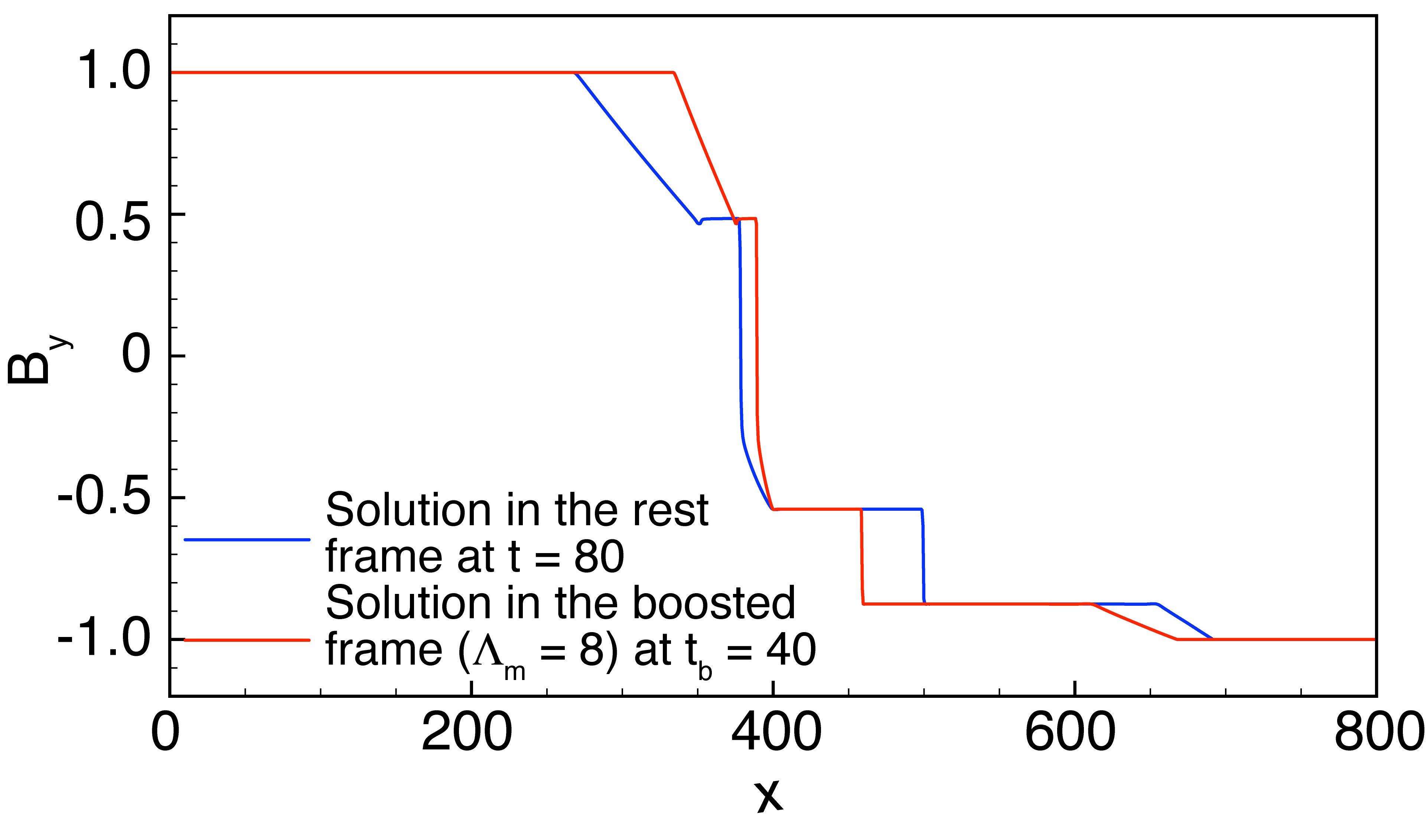}
\includegraphics[width=0.495\linewidth]{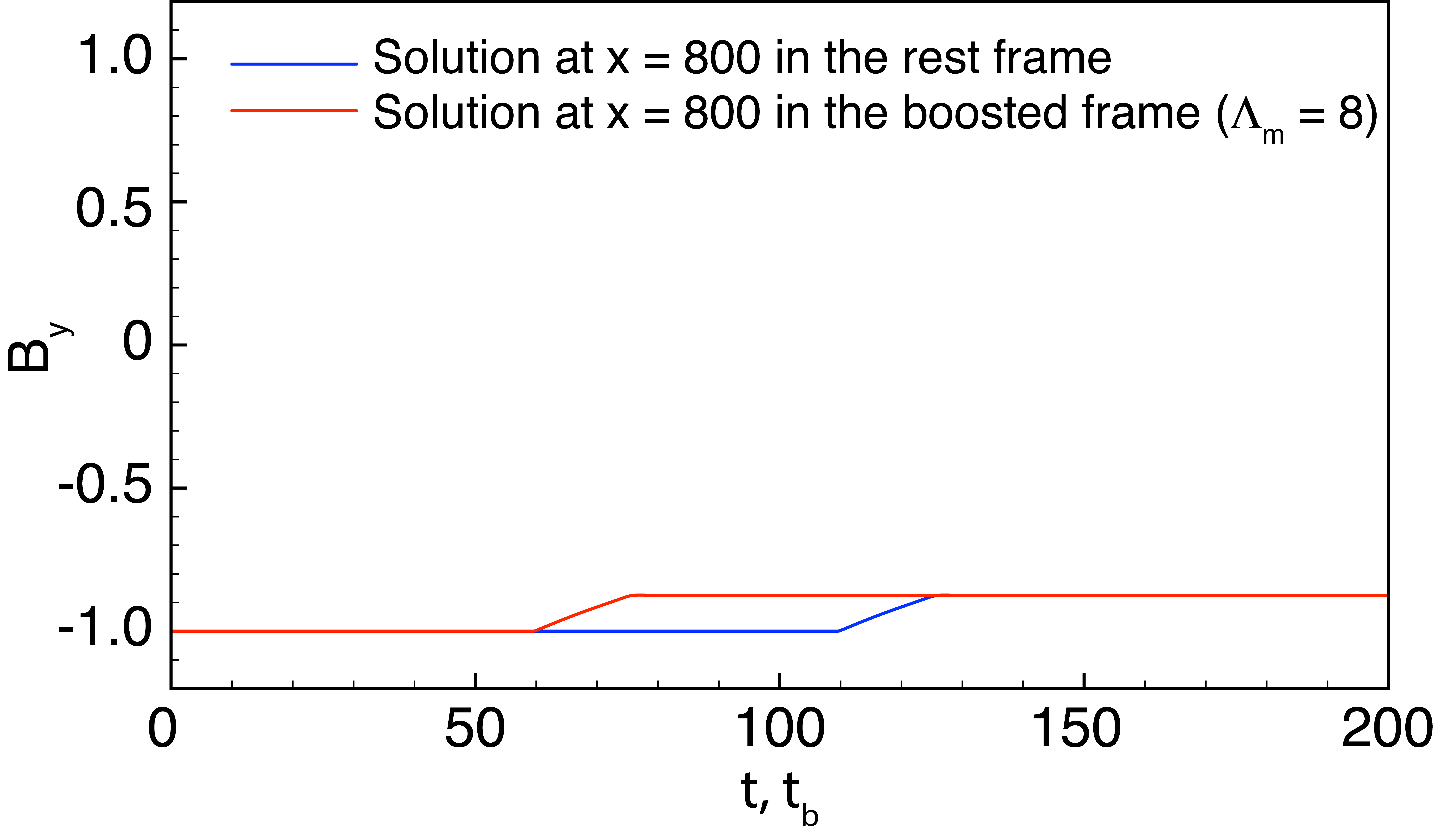}
\caption{Reference solution of the Riemann problem with the initial condition: $\bm{{\cal P}}=(\rho,\bu,\bB,p)=\left(1,0,0,0,\frac34,1,0,1\right)$ at $x<400$; $\bm{{\cal P}}=\left(\frac18,0,0,0,\frac34,-1,0,\frac1{10}\right)$ at $x>400$, for a gas with $\gamma=7/5$. {\bf Left panel:} Blue line shows the solution of MHD equations in the rest frame at the time, $t=80$; red line presents the solution in the boosted frame at time $t_b=40$. {\bf Right panel:} Time profile of density at $x=800$. Red line is the solution in the boosted frame, blue line is the solution in the rest frame. $B_y(t_b)$ predicts the solution in the rest frame, $B_y(t)$ with time offset of $50$.}
\label{fig:mhd_rs}
\end{figure}

As a test problem, we numerically solved the test case \citet{Brio:1988a} \citep[we use the version published by ][]{Cargo1997}, on 800 cells at a spatial resolution of $\Delta x=1$. The initial discontinuity located at $x=400$ separates states $\bm{{\cal P}}=(\rho,\bu,\bB,p)=\left(1,0,0,0,\frac34,1,0,1\right)$ at $x<400$ and $\bm{{\cal P}}=\left(\frac18,0,0,0,\frac34,-1,0,\frac1{10}\right)$ at $x>400$. As an ``exact'' solution, we solve the same problem on 4,000 cells with a resolution of $\mathrm{d}x=0.2$. The solution in the rest frame at time $t=$80 as a function of $0\le x\le 800$ is presented in the left panel of Fig.~\ref{fig:mhd_rs}  by the blue line.

Following Sect.~\ref{Sec:1Dproblem}, we can apply this result as an approximate solution of the Riemann problem in the rest frame, $\bm{{\cal P}}_\mathrm{RS}\left(\frac{x-400}{t}\right)$, in the argument range of $-5\le\frac{x-400}{80}\le5$. In the right panel of Fig.~\ref{fig:mhd_rs} the blue line shows this solution, $\bm{{\cal P}}_\mathrm{RS}\left(\frac{x-400}{t}\right)$, at a fixed location, $x=800$, as a function of time.

The solution of the Riemann problem in the rest frame can also be applied to get an ``exact'' solution in the boosted frame (similarly to Eq.~\ref{eq:Riemannproblem}):
\bea\label{eq:Riemannproblemmhd}
\renewcommand{\arraystretch}{2}
\bm{\mathcal{P}}=\left\{
\begin{array}{ll}
$$\bm{\mathcal{P}}_{\rm RS}\left(\frac{x-400}{t_b}\right)$$ & $$\text{if}\ 0\le x\le 400$$ \\
$$\bm{\mathcal{P}}_{\rm RS}\left(\frac{\frac{x-400}{t_b}}{1+\frac1{\Lambda_m}\frac{x-400}{t_b}}\right)$$ & $$\text{if}\ 400\le x\le 800$$.
\end{array}\right.
\renewcommand{\arraystretch}{1.}
\eea
The reference solution obtained this way in the boosted frame is shown with red lines in Fig.~\ref{fig:mhd_rs} with the choice of $\Lambda_m=8$, corresponding to a time offset, $\Delta t_0=50$, at the right boundary. In the left panel, we show the solution for the magnetic field as a function of $x$ at time $t_b=40$. Although the time interval in the boosted frame is twice shorter than the one used to get the rest frame solution, the distance traveled by the magnetic-field perturbation propagating to the right is close in both cases as a result of the time offset. In the right panel, the red line demonstrates the time evolution of the magnetic perturbation at $x=800$. Comparison with the blue line clearly demonstrates the time offset, $\Delta t_0=50$, \ie in the boosted frame the magnetic perturbation arrives to the right boundary much earlier.

Once the reference solution in the boosted frame is obtained, we can compare it with the numerical simulation results in the boosted frame. In all simulations, we apply the iterative procedure to recover the primitive variables as described in \ref{Subsec:primitive} with the iteration convergence criterion as follows:
\beq\label{eq:convergence}
\|\delta^{h+1}u_\sR\|\le 10^{-6}\Lambda_m.
\eeq

As a cross-check, we also simulated the numerical hydrodynamics test described in \ref{Subsec:numerical1} applying the iterative procedure for the MHD equations to solve the problem where the magnetic field vanishes identically. The results show no noticeable difference from those obtained with the iteration-free procedure to solve for the primitive hydrodynamic variables described in \ref{Subsec:hydro-primitive}.

\begin{figure}[ht!]
\centering
\includegraphics[width=0.495\linewidth]{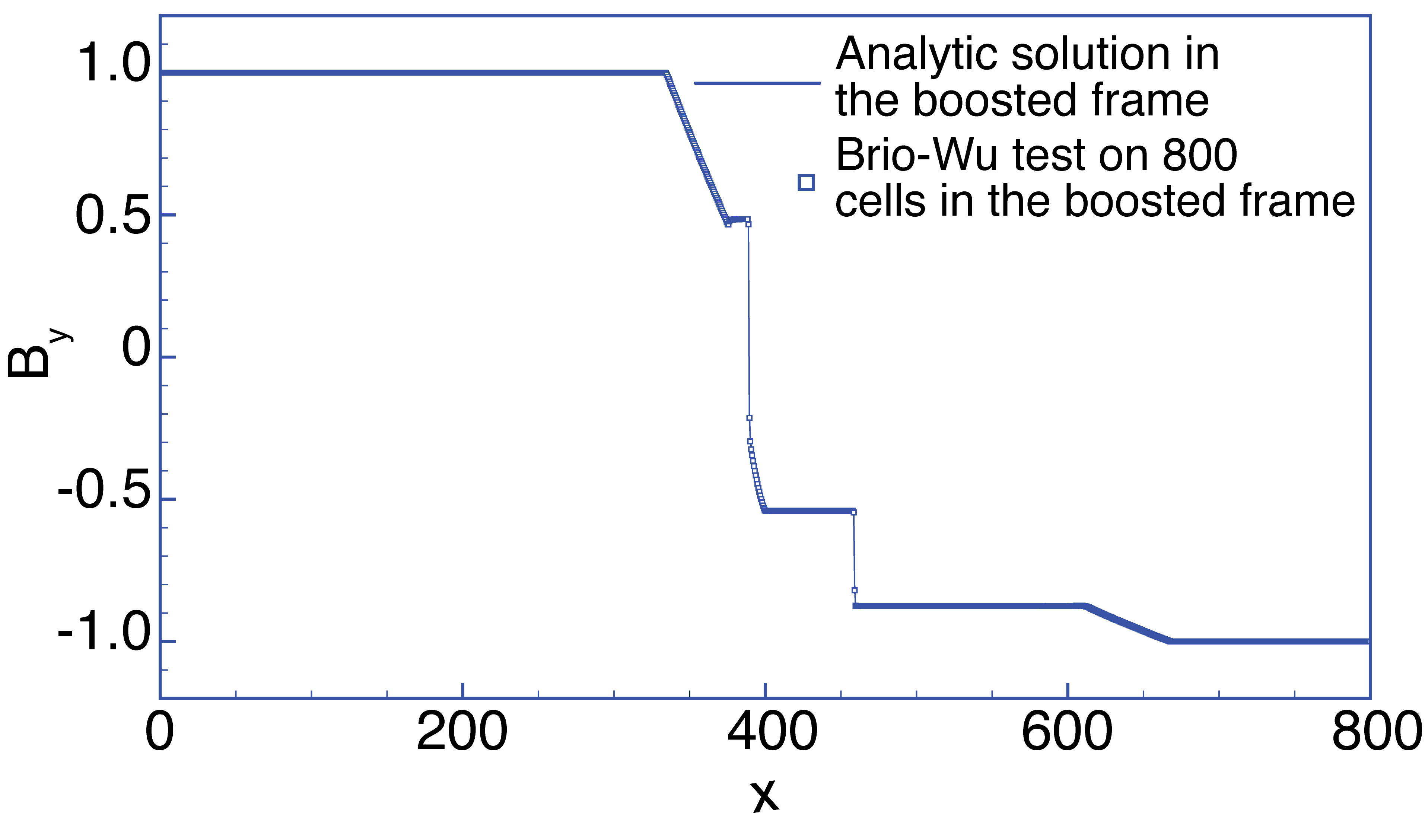}
\includegraphics[width=0.495\linewidth]{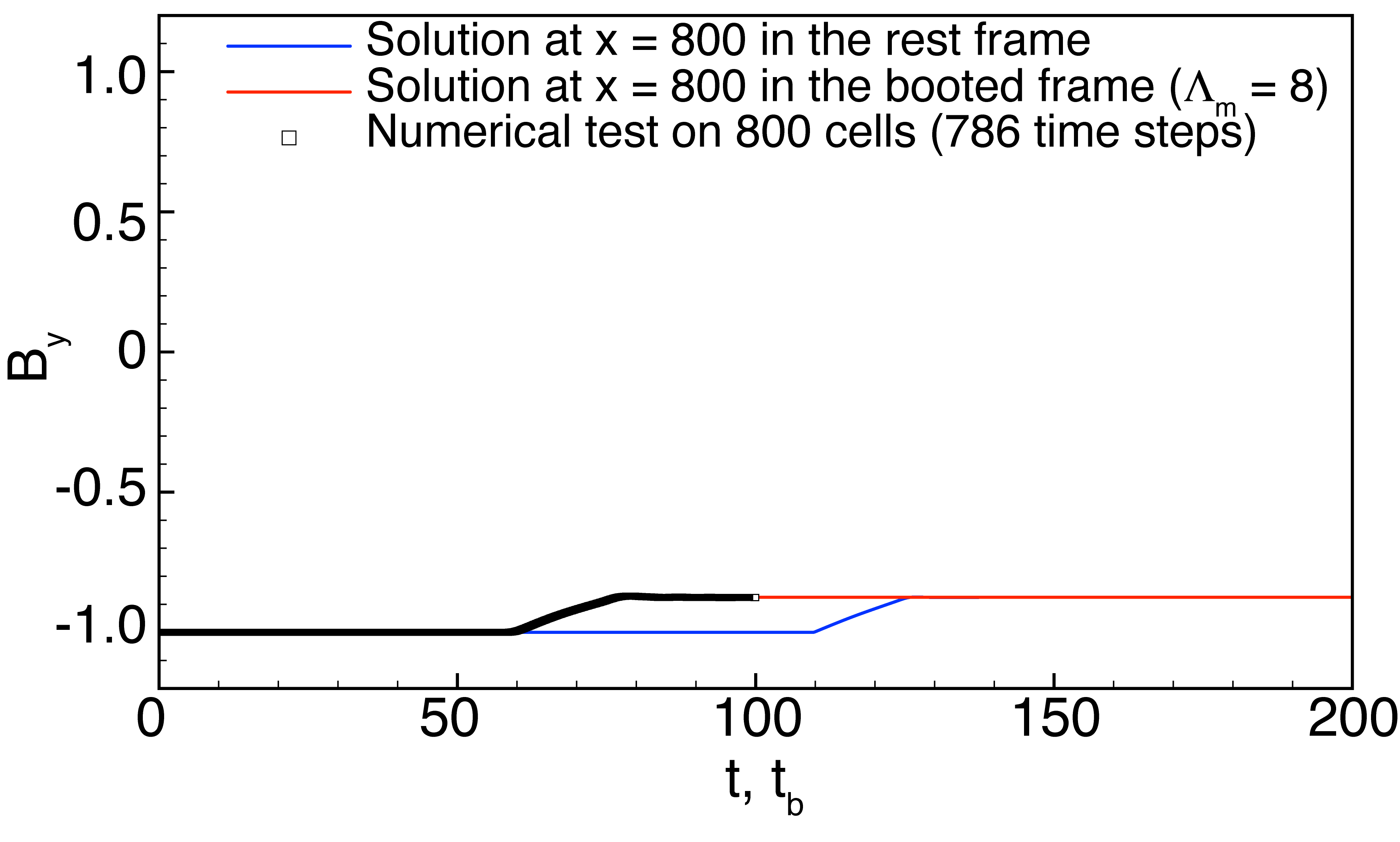}
\caption{Numerical solution of the \citet{Brio:1988a} Riemann problem with the initial condition: $\bm{{\cal P}}=(\rho,\bu,\bB,p)=\left(1,0,0,0,\frac34,1,0,1\right)$ at $x<400$; $\bm{{\cal P}}=\left(\frac18,0,0,0,\frac34,-1,0,\frac1{10}\right)$ at $x>400$, for a gas with $\gamma=7/5$, in the boosted frame with $\Lambda_m=8$. The  {\bf Left panel:} magnetic field as a function of $x$ at time $t_b=40$: the solid line shows the exact solution (also shown with the red line in Fig.~\ref{fig:mhd_rs}); while symbols represent the numerical test solution on $800$ cells.  {\bf Right panel:} Time profile of the magnetic field at $x=800$. Red line is the exact solution in the boosted frame, blue line is the exact solution in the rest frame (both lines are the same as those in Fig.~\ref{fig:mhd_rs}); symbols show the numerical test result for an interval, $(0,100)$ of $t_b$ (at 786 time steps), with 800 cells per an interval, $(0,800)$, of x.}
\label{fig:testmhd}
\end{figure}

As an MHD test, we numerically solved the \citet{Brio:1988a} problem in the boosted frame with $\Lambda_m=8$ on 800 cells. We applied the \textit{artificial wind} \citep{Sokolov2002} numerical flux combined with the limited reconstruction procedure to interpolate primitive variables on faces. A two-stage Runge-Kutta scheme allowed us to achieve the second order of accuracy in time. In the left panel of Fig.~\ref{fig:testmhd} we show the numerical result for the magnetic field distribution over $x$ at time $t_b=40$. The agreement with the reference solution in the boosted frame (shown in the left panel of Fig.~\ref{fig:mhd_rs}) is remarkable. 

To obtain the test result for the temporal evolution in a fixed location, $x=800$, we extend the computational domain by 400 cells on both sides, to reduce the effect of the boundaries of the computational domain on the numerical solution. The numerical result for the magnetic field at $x=800$ as a function of time is shown in the right panel of Fig.~\ref{fig:testmhd} by symbols. For comparison, the blue line and the red line show the exact solution in the rest frame and in the boosted frame correspondingly (both lines are the same as those in Fig.~\ref{fig:mhd_rs}). The numerical result in the boosted frame both agrees with the exact solution in the boosted frame and perfectly ``predicts'' the solution in the rest frame with the time offset of $(800-400)/\Lambda_m=50$.

\section{Future Work on Numerical Algorithms}
\label{Sec:future}
In the appendices of this paper we discussed the major issues to be solved in applying our proposed approach to a simplified model of the solar wind. However, the choice of a more sophisticated model will require more analytic work. Specifically, the ``non-conservative" model, in which instead of the full energy conservation law the non-conservative internal energy equation is used, which in the rest frame reads::
\beq
\frac{\partial}{\partial t}\frac{p}{\gamma - 1}+\nabla\cdot\left(\frac{p\bu}{\gamma-1}\right)+p\nabla\cdot\bu=0.
\eeq
In this equation the transformation of the velocity divergence should be applied in the boosted frame similarly to how the magnetic field divergence is treated in Sect.~\ref{Sec:mhd} and \ref{Sec:numericalmhd}: $\nabla\cdot\bu\rightarrow \nabla\cdot\bu-\frac1{\Lambda_m}\frac{\partial u_R}{\partial t_b}$. If the electron temperature is not assumed to be equal to the ion one, this kind of equation also needs to be solved for the electron pressure. In this case electron internal energy and pressure contribute to the total energy and pressure, respectively, while the electron pressure needs to be solved from the following equation:
\beq\label{eq:electronpressure}
\frac{\partial}{\partial t}\frac{p_e}{\gamma_e - 1}+\nabla\cdot\left(\frac{p_e\bu}{\gamma_e-1}\right)+p_e\nabla\cdot\bu=0.
\eeq
\end{document}